\documentclass[twocolumn,prd,aps,showpacs,amsmath,amssymb]{revtex4-1}
\usepackage{graphics}
\usepackage{graphicx}
\usepackage{dcolumn}
\usepackage{bm}
\usepackage{mathrsfs}
\usepackage{pstricks}
\usepackage{color}
\usepackage{slashed}
\usepackage{amsmath}
\usepackage{epsfig}
\usepackage{amsfonts}
\usepackage{amssymb}

\def\beq{\begin{equation}}
\def\eeq{\end{equation}}
\def\bea{\begin{eqnarray}}
\def\eea{\end{eqnarray}}
\def\nn{\nonumber}

\def\Re{\textrm{Re}}

\begin{document}

\title{Radiative processes of two entangled atoms outside a Schwarzschild black hole}
\author{G. Menezes}
\email{gabrielmenezes@ufrrj.br}
\affiliation{Grupo de F\'isica Te\'orica e Matem\'atica F\'isica, Departamento de F\'isica, Universidade Federal Rural do Rio de Janeiro, 23897-000 Serop\'edica, RJ, Brazil}
%

\begin{abstract}
We consider radiative processes of a quantum system composed by two identical two-level atoms in a black-hole background. We assume that these identical two-level atoms are placed at fixed radial distances outside a Schwarzschild black hole and interacting with a quantum electromagnetic field prepared in one of the usual vacuum states, namely the Boulware, Unruh or the Hartle-Hawking vacuum states. We study the structure of the rate of variation of the atomic energy. The intention is to identify in a quantitative way the contributions of vacuum fluctuations and radiation reaction to the entanglement generation between the atoms as well as the degradation of entangled states in the presence of an event horizon. We find that for a finite observation time the atoms can become entangled for the case of the field in the Boulware vacuum state, even if they are initially prepared in a separable state. In addition, the rate of variation of atomic energy is not well behaved at the event horizon due to the behavior of the proper accelerations of the atoms. We show that the thermal nature of the Hartle-Hawking and Unruh vacuum state allows the atoms to get entangled even if they were initially prepared in the separable ground state.  
\end{abstract}

\pacs{03.65.Ud, 03.67.-a, 04.62.+v, 04.70.Dy, 42.50.Lc, 97.60.Lf}

\maketitle

\section{Introduction}
\label{intro}

Quantum entanglement is the essencial feature underlying quantum information, cryptography and quantum computation~\cite{1,haroche}. Systems of two-level atoms interacting with a bosonic field have been one of the leading prototypes in the investigations concerning entangled states~\cite{2,3,4,5,ved}. In turn, radiative processes of entangled states have been substantially considered in the literature~\cite{aga,rep}. Here we quote Ref.~\cite{yang} in which the authors investigate the properties of emission from two entangled atoms coupled with an electromagnetic field in unbounded space. In addition, in Ref.~\cite{eberly} the authors demonstrate for spontaneous emission processes how nonlocal disentanglement times can be shorter than local decoherence times for arbitrary entangled states. See also Ref.~\cite{fi}.

The field of relativistic quantum information has emerged in recent years as an active research program connecting concepts from gravitational physics and quantum computing. With this respect, several important works were developed~\cite{fue,bene,als3,ber,mira1,mira2}. We also quote  references~\cite{rez1,rez2,braun,mass,fran,LinHu2009,LinHu2010} which establish important results concerning entanglement generation between two localized causally disconnected atoms. On the other hand, many investigations were also implemented on a curved background. For instance, it was shown in reference~\cite{ver} that an expanding space-time acts as a decohering agent which forces the entanglement of the vacuum to greatly decrease due to the effects of the Gibbons-Hawking temperature~\cite{gibbons}. Another example, of immense current interest, is related with investigations of quantum entanglement in a Schwarzschild space-time which was undertaken in Ref.~\cite{mar}. This was also the subject of study by the authors in Ref.~\cite{china4}. In such a reference entanglement was considered in the framework of open quantum systems. In the presence of a weak gravitational field, the authors in~\cite{kempf} has given manifest evidence that the amount of entanglement that Unruh-DeWitt detectors can extract from the vacuum can be increased. Similar results were found in references~\cite{ben1,ben2,china3}. For a review of recent results regarding entanglement in curved space-times we refer the reader the work~\cite{mar2} and references cited therein. The point that should be emphasized is that many of such studies seem to imply the importance of considering the observer-dependency property of quantum entanglement~\cite{als4}. Hence a detailed understanding of such phenomena is mandatory in investigations concerning quantum information processes in the presence of a gravitational field or, more specifically, near an event horizon.

The aim of the present paper is to contribute to the investigations of relativistic quantum information theory in the light of an alternative perspective. Most of the investigations aforementioned were implemented in a framework of open quantum systems, or by employing time-dependent perturbation theory or similar techniques. The heuristic picture raised in such methods is that the generation (or degradation) of entanglement between two-level atoms is triggered by the vacuum fluctuations of the quantum field. With this respect, in a recent work radiative processes of entangled atoms interacting with a massless scalar field prepared in the vacuum state in the presence of boundaries were considered~\cite{juan}. Nevertheless, when discussing stimulated emission and absorption which have equal Einstein $B$ coefficients, it is not clear whether vacuum fluctuations always act as the only source of (or degradation of) entanglement. This is a consequence of the fact that it is possible to interpret spontaneous decay as a radiation-reaction effect~\cite{ake}. As carefully demonstrated by Milonni, both effects, vacuum fluctuations and radiation reaction, depend on the ordering chosen for commuting atomic and field operators~\cite{mil1}. Following such debates recently quantum entanglement between inertial atoms~\cite{ng1} and uniformly accelerated atoms~\cite{ng2} coupled with an electromagnetic field was discussed in the framework developed by Dalibard, Dupont-Roc, and Cohen-Tannoudji (DDC)~\cite{cohen2,cohen3}. The results for uniformly accelerated atoms compare with the situation in which two atoms at rest are coupled individually to two spatially separated cavities at different temperatures, recovering, in some sense, the outcomes described in~\cite{eberly}. In addition, for equal accelerations it was obtained that one of the maximally entangled antisymmetric Bell state is a decoherence-free state. 

We remark that the DDC formalism was also successfully implemented in many interesting physical situations~\cite{aud1,aud2,china1,china2,rizz}, including quantum fields in curved space-time~\cite{china5,china6,china}. For uniformly accelerated atoms, such a method quantitatively motivates the scenario presented in~\cite{sciama}. On the other hand, in the investigations concerning quantum entanglement, the DDC formalism has proved to be a pivotal treatment in order to better understand the structure responsible for supporting entanglement in radiative processes involving atoms, as demonstrated in Refs.~\cite{ng1,ng2}. Specifically, in such references it was shown how the rate of variation of atomic energy evaluated within the DDC approach can be an useful quantity in order to signalize the emergence of quantum entanglement. This is the idea we intend to continue to explore further in this work by considering the resonant interaction between atoms in a Schwarzschild space-time.

Even though close to event horizon the Schwarzschild metric takes the form of the Rindler line element, there are important distinctions between an event horizon in Schwarzschild space-time and an acceleration horizon in Rindler space-time. In the present paper we propose to generalize the results of Refs.~\cite{ng1,ng2} for the case of identical two-level atoms in a Schwarzschild space-time. Being more specific, we intend to investigate these atoms coupled with quantum electromagnetic fluctuations in the Boulware, Unruh and Hartle-Hawking vacuum states. We use the approach above discussed which allows an easy comparison of quantum mechanical and classical concepts. The organization of the paper is as follows. In Section~\ref{model} we discuss the identification of vacuum fluctuations and radiation reaction effect in the situation of interest. In Section~\ref{atom} we calculate the rates of variation of atomic energy with finite observation time intervals for atoms placed at fixed radial distances outside a Schwarzschild black hole. Conclusions and final remarks are given in Section~\ref{conclude}. In the Appendices we briefly digress on the correlation functions of electromagnetic field in Schwarzschild space-time. We also discuss asymptotic evaluation of mode sums which will be important in what follows. In this paper we use units such that $\hbar = c = G = k_B = 1$. 

\section{The coupling of atoms with electromagnetic fields in black-hole space-time}
\label{model}

Let us suppose the case of two identical two-level atoms interacting with a common electromagnetic field. In this paper we work in the multipolar coupling scheme which means that all interactions are realized through the quantum electromagnetic fields. This formalism is suitable for describing retarded dipole-dipole interactions between the atoms. In general, the atoms will be moving along different world lines, so there will be two different proper times parameterizing each of these curves. We are working in a four-dimensional Schwarzschild space-time, which is described by the line element:
\bea
ds^2 &=& \left(1 - \frac{2M}{r}\right)\,dt^2 -\left(1 - \frac{2M}{r}\right)^{-1}dr^2 
\nn\\
&-&\, r^2 (d\theta^2 + \sin^2\theta d\phi^2), 
\label{schw}
\eea
which is the vacuum solution to the Einstein field equations that describes the gravitational field outside a spherically symmetric body of mass $M$ (we employ the convention that the Minkowski metric is given by $\eta_{\alpha\beta} = -1, \alpha=\beta=1,2,3$, $\eta_{\alpha\beta} = 1, \alpha=\beta=0$ and $\eta_{\alpha\beta} = 0,\alpha \neq \beta$). The collapse of an electrically neutral star endowed with spherical symmetry produces a spherical black hole of mass $M$ with external gravitational field described by the Schwarzschild line element~(\ref{schw}). The surface of the balck hole, i.e., the event horizon, is located at $r = 2M$, the position where the Schwarzschild coordiantes become singular. Only the region on and outside the black hole's surface, $r \geq 2M$, is relevant to external observers. Events inside the horizon can never influence the exterior, at least in the classical regime. An interesting discussion on Schwarzschild's original solution can be found in reference~\cite{antoci}.

Here we are interested in the situation of  black-hole background geometry. Being more specific, we are prompted to dispense entirely with the spherically symmetric body and examine the quantum field theory for the electromagnetic field on the maximally extended manifold which is everywhere a solution of the vacuum Einstein equation. This is obtained from equation~(\ref{schw}) by replacing the coordinates $(t, r)$ by the so-called Kruskal-Szekeres coordinates $(v, u)$. For an extensive discussion, see references~\cite{wheeler,birrel}. The Schwarzschild geometry consists of four different regions, see figure ($31.3$) of Ref.~\cite{wheeler}. Regions I and III portray two distinct asymptotically flat universes with $r > 2M$; in fact, in region $\textrm{III}$ the coordinate time $t$ runs backwards with respect to region $\textrm{I}$. Regions II and IV are also time-reversed regions in which physical singularities ($r = 0$) evolve. In the Kruskal-Szekeres coordinates one can show that the metric is perfectly well defined and non-singular at the event horizon. In addition, such transformations and the metric~(\ref{schw}) make clear that near the event horizon the line element approaches the form of the Rindler line element. Therefore, for $r \approx 2M$ the Schwarzschild coordinates $t$ and $r$ behave as Rindler space-time coordinates.

In this paper we propose to identify quantitatively the contributions of quantum field vacuum fluctuations and radiation reaction to the entanglement dynamics of atoms in black-hole space-time. With this respect, one must consider the Heisenberg picture. We consider both atoms moving along different stationary trajectories $x^{\mu}(\tau_i) = (t(\tau_i),{\bf x}(\tau_i))$, where $\tau_i$ denotes the proper time of the atom $i$.  Due to this fact, in what follows we describe the time evolution with respect to the Schwarzschild coordinate time $t$ which, because of~(\ref{schw}), has a functional relation with each of the proper times of the atoms. 

We suppose that the two-level atoms are placed at fixed radial distances outside the black hole. The stationary trajectory condition guarantees the existence of stationary states. Within the multipolar-coupling scheme the purely atomic part of the total Hamiltonian describes the free atomic Hamiltonian. A brief and important comment is in order. It is known that the presence of gravitational fields affects the Coulomb interaction between charges within the atoms as well as dipole energies~\cite{ka,lam}. In addition, van der Waals forces are modified by gravity~\cite{pinto}. As a first approximation, we shall consider that the coupling between the atoms and the gravitational field is sufficiently weak. Hence we take the free atomic Hamiltonian as having the same functional form as in the absence of gravitation. In this context, the Hamiltonian of this atomic system can then be written as
\bea
H_A(t) &=& \frac{\omega_0}{2}\biggl[\left(\sigma_{1}^z(\tau_1(t))\otimes\hat{1}\right)\,\frac{d\tau_1}{dt} 
\nn\\
&+&\,\left(\hat{1}\otimes \sigma_{2}^z(\tau_2(t))\right)\,\frac{d\tau_2}{dt}\biggr],
\label{ha}
\eea
where $d\tau/dt  = \sqrt{g_{00}} = (1 - 2M/r)^{1/2}$ and $\sigma_a^z = | e_a \rangle\langle e_a | - | g_a \rangle\langle g_a |$, $a = 1, 2$. Here $|g_1\rangle$, $|g_2\rangle$ and $|e_1\rangle$, $|e_2\rangle$ denote the ground and excited states of isolated  atoms, respectively. One has that the space of the two-atom system is spanned by four product states with respective eigenenergies
\bea
&& E_{gg} = -\omega_0\,\,\,\,|gg\rangle = |g_1\rangle|g_2\rangle, 
\nn\\
&& E_{ge}= 0\,\,\,\,|ge\rangle = |g_1\rangle|e_2\rangle,
\nn\\
&&E_{eg} = 0\,\,\,\,|eg\rangle = |e_1\rangle|g_2\rangle,
\nn\\
&&E_{ee} = \omega_0\,\,\,\,|ee\rangle = |e_1\rangle|e_2\rangle.
\label{sta}
\eea
Here we consider that the two-atom system is coupled with an electromagnetic field. The Hamiltonian $H_F(t)$ of the free electromagnetic field can be obtained in the usual way from equation~(\ref{ac-field}), see Appendix~\ref{A}. In this way, one has that
\beq
H_F(t) = \sum_{{\bf k}}\, \omega_{{\bf k}}\,a^{\dagger}_{{\bf k}}(t)a_{{\bf k}}(t),
\label{hf}
\eeq
where $a^{\dagger}_{{\bf k},\lambda}, a_{{\bf k},\lambda}$ are the usual creation and annihilation operators of the electromagnetic field and we have neglected the zero-point energy. In addition, ${\bf k}$ labels the wave vector and polarization of the field modes. Furthermore, we also assume that the presence of a gravitational field does not affect substantially the physical consequences in considering the interaction between the atoms and the fields. Hence in the multipolar coupling scheme and using the electric-dipole approximation one has that the Hamiltonian which describes the interaction between the atoms and the field is given by
\bea
H_{I}(t) &=& - \boldsymbol\mu_1(\tau_1(t)) \cdot {\bf E}(x_1(\tau_1(t)))\,\frac{d\tau_1}{dt} 
\nn\\
&-&\, \boldsymbol\mu_2(\tau_2(t)) \cdot 
{\bf E}(x_2(\tau_2(t)))\,\frac{d\tau_2}{dt}
\label{hi}
\eea
where $\boldsymbol\mu_i$ ($i=1,2$) is the electric dipole moment operator for the $i$-th atom. The electric field above is the measured electric field defined through the measured force it exerts on the atoms. The dipole moment operator is given by
\beq
\boldsymbol\mu_i(\tau_i) = \boldsymbol\mu\left[\sigma_{i}^{+}(\tau_i) + \sigma_{i}^{-}(\tau_i)\right],
\label{dip}
\eeq
where we have assumed that the dipole matrix elements $\langle g_i |\boldsymbol\mu_i| e_i \rangle$ are real and we denote them by $\boldsymbol\mu$ since they are independent of the index $i$ (identical and similarly oriented atoms). In the above we have defined the raising and lowering operators as $\sigma_{i}^{+} = | e_i \rangle\langle g_i|$ and $\sigma_{i}^{-} = | g_i \rangle\langle e_i|$, respectively. Incidentally, suppose that our atoms are spinless one-electron systems. Hence $\boldsymbol\mu_a = e\,\hat{{\bf r}}_a$, where $e$ is the electron charge and $\hat{{\bf r}}_a$ is the position operator of the atom $a$.

The Heisenberg equations of motion for the dynamical variables of the atom and the field with respect to $t$ can be derived from the total Hamiltonian $ H(t) = H_A(t) + H_F(t) + H_I(t)$. After establishing the equations of motion, in order to solve them one usually separates the solutions in two parts, namely: The free part, which is independent of the presence of a coupling between atoms and fields; and the source part, which is caused by the interaction between atoms and fields. That is, for atomic and field operators, respectively: $\sigma_a^z(\tau_a(t)) = \sigma_a^{z,f}(\tau_a(t)) + \sigma_a^{z,s}(\tau_a(t))$ and also $a_{{\bf k}}(t) = a^{f}_{{\bf k}}(t) + a^{s}_{{\bf k}}(t)$. Since one can construct from the annihilation and creation field operators the free and source part of the quantum electric field, one can also write ${\bf E}(t) = {\bf E}^{f}(t) + {\bf E}^{s}(t)$. As extensively discussed in Refs.~\cite{cohen2,cohen3,aud1}, this calculation produces an ambiguity of operator ordering. In summary, this implies that one must choose an operator ordering when discussing the effects of ${\bf E}^{f}$ and ${\bf E}^s$ separately. This is the root of the feature already discussed in the Introduction by which the effects of vacuum fluctuations (which is caused by ${\bf E}^{f}$) and radiation reaction (which is originated from ${\bf E}^s$) depend on the ordering chosen for commuting atomic and field operators. Nonetheless, here we adopt a particular prescription which enables to interpret the effects of such phenomena as independent physical processes~\cite{cohen2,cohen3,aud1}. This is essentially the DDC formalism mentioned above.

We do not intend to give a thorough treatment of the DDC formalism here, since this approach has been analyzed in detail in many works. The reader may benefit from reading the several expositions we have already quoted above, specially reference~\cite{aud1} which, to the best of this author's knowledge, was one of the first works to discuss the Unruh effect~\cite{birrel,nami,mat} within such a framework. Therefore, we only expound the main results. The idea is to evaluate $d H_A/ dt$, where $H_A$ is given by equation~(\ref{ha}), and consider only the part which is due to the interaction with the field; afterwards one extracts from the remaining quantity the contributions of vacuum fluctuations and radiation reaction and then one takes the expectation value of the resulting quantities. The latter consists of two different operations: first we consider an averaging over the field degrees of freedom (obtained by taking vacuum expectation values); subsequently one takes the expectation value of the associated expressions in an atomic state $|\nu\rangle$, with energy $\nu$. Such a state is usually one of the product states given by equation~(\ref{sta}) but it can be any given state. For the purposes of studying entanglement, one can conveniently take $| \nu \rangle$ as a generic entangled state. For instance, consider the entangled states:
\beq
|\Omega^{\pm}\rangle = c_1\,|g_1\rangle|e_2\rangle \pm c_2\,|e_1\rangle|g_2\rangle,
\label{bell}
\eeq
where $c_1, c_2$ are complex numbers. Note that $|\Omega^{\pm}\rangle$ are eigenstates of the atomic Hamiltonian $H_A$. Here we will be particularly interested in the situation where $c_1 = c_2 = 1/\sqrt{2}$. Such states constitute familiar examples of maximally entangled Bell states. The other Bell states are given by:
\beq
|\Phi^{\pm}\rangle = \frac{1}{\sqrt{2}}\left(|g_1\rangle|g_2\rangle \pm |e_1\rangle|e_2\rangle\right).
\eeq
The Bell states form an alternative basis of the two-qubit Hilbert space. They play a fundamental role in Bell measurements and they are also known as the four maximally entangled two-qubit Bell states.

Coming back to our problem, let us present the contributions of vacuum fluctuations and radiation reaction in the evolution
of the atoms' energies. Proceeding with a usual perturbative expansion and taking into account only terms up to order $\mu^2$, the vacuum-fluctuation contribution reads
\begin{widetext}
\beq
\Biggl\langle \frac{d H_A}{dt} \Biggr\rangle_{VF} = \frac{i}{2}\int_{t_0}^{t}dt' \,\sum_{a,b = 1}^{2}\,\frac{d\tau_a}{dt}\frac{d\tau_b'}{dt'}\,D_{ij}(x_a(\tau_a(t)),x_b(\tau_b'(t')))\frac{\partial}{\partial\tau_a}\Delta^{ij}_{ab}(\tau_a(t),\tau_b'(t')),
\label{vfha3}
\eeq
where the notation $\langle (\cdots) \rangle = \langle 0,\nu |(\cdots)| 0, \nu \rangle$ has been employed ($| 0\rangle$ is the vacuum state of the field, to be discussed below). In the above:
\beq
\Delta^{ij}_{ab}(\tau_a(t),\tau_b'(t')) = \langle \nu | [\mu^{i,f}_{a}(\tau_a(t)),\mu^{j,f}_{b}(\tau_b'(t'))]| \nu \rangle,\,\,\,a,b = 1,2,
\label{susa}
\eeq
is the linear susceptibility of the two-atom system in the state $|\nu\rangle$ and 
\beq
D_{ij}(x_a(\tau_a(t)),x_b(\tau_b'(t'))) = \langle 0 |\{E^{f}_{i}(x_a(\tau_a(t))),E^{f}_{j}(x_b(\tau_b'(t')))\}| 0 \rangle,
\eeq
$a, b =1,2$ is the Hadamard's elementary function. On the other hand, for the radiation-reaction contribution, one has:
\beq
\Biggl\langle \frac{d H_A}{dt} \Biggr\rangle_{RR} = \frac{i}{2}\int_{t_0}^{t}dt' \,\sum_{a,b=1}^{2}\,\frac{d\tau_a}{dt}\frac{d\tau_b'}{dt'}\,\Delta_{ij}(x_a(\tau_a(t)),x_b(\tau_b'(t')))\frac{\partial}{\partial\tau_a} D^{ij}_{ab}(\tau_a(t),\tau_b'(t')),
\label{rrha3}
\eeq
where
\beq
D^{ij}_{ab}(\tau_a(t),\tau_b'(t')) = \langle \nu | \{\mu^{i,f}_{a}(\tau_a(t)),\mu^{j,f}_{b}(\tau_b'(t'))\}| \nu \rangle,\,\,\,a,b = 1,2,
\label{cora}
\eeq
is the symmetric correlation function of the two-atom system in the state $|\nu\rangle$ and 
\beq
\Delta_{ij}(x_a(\tau_a(t)),x_b(\tau_b'(t'))) = \langle 0 |[E^{f}_{i}(x_a(\tau_a(t))),E^{f}_{j}(x_b(\tau_b'(t')))]| 0 \rangle,
\eeq
\end{widetext}
$a, b =1,2$ is the Pauli-Jordan function. We see from equations~(\ref{vfha3}) and~(\ref{rrha3}) that one can identify two distinct contributions. One is due to the existence itself of the atoms and it is independent of any interaction whatsoever. The other is related with the emergence of cross correlations between the atoms mediated by the field. Likewise, observe that such a formalism enables one to discuss the interplay between vacuum fluctuations and radiation reaction in the generation or degradation of entanglement between atoms.

As emphasized in many texts, $\Delta^{ij}_{ab}$ and $D^{ij}_{ab}$ characterize only the two-atom system itself. The explicit forms of such quantities are given by
\bea
\Delta^{ij}_{ab}(t,t') &=& \sum_{\nu'} \biggl[{\cal U}_{ab}^{ij}(\nu, \nu')\,e^{i\Delta\nu(\tau_a(t) - \tau_b(t'))} 
\nn\\
&-&\,{\cal U}_{ba}^{ji}(\nu, \nu')\,e^{-i\Delta\nu(\tau_a(t) - \tau_b(t'))}\biggr],
\label{susa1}
\eea
and
\bea
D^{ij}_{ab}(t,t') &=& \sum_{\nu'} \biggl[{\cal U}_{ab}^{ij}(\nu, \nu')\,e^{i\Delta\nu(\tau_a(t) - \tau_b(t'))} 
\nn\\
&+&\, {\cal U}_{ba}^{ji}(\nu, \nu')\,e^{-i\Delta\nu(\tau_a(t) - \tau_b(t'))}\biggr],
\label{cora1}
\eea
where $\Delta \nu = \nu - \nu'$ and we have conveniently introduced a suitable generalized atomic transition dipole moment ${\cal U}_{ab}^{ij}(\nu, \nu')$ defined as
\beq
{\cal U}_{ab}^{ij}(\nu, \nu') =  \langle \nu |\mu^{i,f}_{a}(0)| \nu' \rangle\langle \nu' |\mu^{j,f}_{b}(0)| \nu \rangle.
\label{aa}
\eeq
Finally, observe that from~(\ref{schw}) one can easily perform a change of variables in equations~(\ref{vfha3}) and~(\ref{rrha3}) in order to describe the time evolution in terms of one of the proper times of the atoms. In fact, the use of the proper time is the customary procedure since it is the quantity directly measurable by the clocks of the observers. However, as remarked above, here we adopt an alternative method in which we use the Schwarzschild coordinate time as the parameter that describes the time evolution of the system.

Now we are ready to characterize the entanglement generation (or degradation) between atoms as transitions between particular stationary states of the atomic Hamiltonian. The rate of variation of the atomic energy clearly identifies the permissible transitions between states and depending on the nature of the initial and final states one may plainly perceive the constitution (or destruction) of an entangled state. In particular, as discussed above, within the DDC formalism we can study how the interplay between vacuum fluctuations and radiation reaction significantly influences the occurrence of these phenomena. Hence we propose to investigate the creation of entanglement as well as how entangled states reduce to separable states. For instance, assume that the atoms were initially prepared in an entangled state, that is $|\nu\rangle = |\Omega^{\pm}\rangle$. Hence the only allowed transitions are $|\Omega^{\pm}\rangle \to |gg\rangle$, with $\Delta \nu = \nu - \nu' = \omega_0 > 0$ and $|\Omega^{\pm}\rangle \to |ee\rangle$, with $\Delta \nu = \nu - \nu' = - \omega_0 < 0$. In other words, the rate of variation of atomic energy should indicate the probability for the transitions $|\Omega^{\pm}\rangle \to |gg\rangle$ or $|\Omega^{\pm}\rangle \to |ee\rangle$ by displaying a nonzero value. On the other hand, suppose that the atoms were initially prepared in the atomic ground state ($|\nu\rangle = |gg\rangle$). The transition rates to one of the entangled states $|\Omega^{\pm}\rangle$ are nonvanishing, with the energy gap $\Delta \nu = - \omega_0 < 0$. In all such transitions, the non-zero matrix elements are given by, with $c_1 = c_2 = 1/\sqrt{2}$:
\bea
{\cal U}^{ij}_{11}(\nu, \nu') &=& \frac{\mu^{i}\mu^{j}}{2}
\nn\\
{\cal U}^{ij}_{22}(\nu, \nu') &=& \frac{\mu^{i}\mu^{j}}{2}
\nn\\
{\cal U}^{ij}_{12}(\nu, \nu') &=& {\cal U}_{21}(\nu, \nu') = \pm\, \frac{\mu^{i}\mu^{j}}{2},
\label{mel}
\eea
where $\nu$ stands for the ground state $|gg\rangle$ (or the excited state $|ee\rangle$) and $\nu'$ stands for the entangled states $|\Omega^{\pm}\rangle$, or vice-versa.

In the next Section we will consider in detail the rate of variation of atomic energy for atoms at rest in various important physical situations.\\

\section{Rate of variation of the atomic energy in vacuum}
\label{atom}

As discussed, we consider our two-atom system in a situation where the atoms are placed at fixed radial distances with the world lines given respectively by $x^{\mu}(\tau_i) = (\tau_i/\sqrt{g_{00(r_i)}} , r_i, \theta_i, \phi_i)$, $i = 1, 2$ and $g_{00}(r) = 1 - 2M/r$. Let us investigate the rate of change of the  atomic energy of the two-atom system for each one of the possible vacua discussed in Appendix~\ref{A}. We consider the transitions discussed at the end of the previous Section. For simplicity, we assume that the atoms are polarized along the radial direction defined by their positions relative to the black-hole space-time rotational Killing vector fields. This means that we do not need to calculate the contributions associated with the polarizations in the $\theta$− and $\phi$− directions and the only field correlation functions that we should evaluate are the ones associated with the radial component of the electric field. This is extensively discussed in Appendix~\ref{A}. In the course of the calculations, one will typically deal with asymptotic estimation of mode sums. In the cases of interest this is substantially discussed in Appendix~\ref{B}. 

\subsection{The Boulware vacuum}

The Boulware vacuum has a close similarity to the innermost concept of an empty state at large radii, but it has pathological behavior at the horizon: the renormalized expectation value of the stress tensor, in a freely falling frame, diverges as $r \to 2M$~\cite{sciama}. The Boulware vacuum is the appropriate choice of vacuum state for quantum fields in the vicinity of an isolated, cold neutron star.

We now proceed to calculate the rate of variation of atomic energy in the Boulware vacuum state. From the results derived in the Appendix~\ref{A}, one may compute all the relevant correlation functions of the electric field which appears in equations~(\ref{vfha3}) and~(\ref{rrha3}). The associated Hadamard's elementary functions are given by
\begin{widetext}
\bea
D^{B}_{rr}(x_i(t),x_j(t')) &=& \frac{1}{16\pi^2}\sum_{l = 1}^{\infty}(2l +1)P_{l}(\hat{r}_i \cdot \hat{r}_j)
\int_{0}^{\infty}\,d\omega\,\omega\,\Biggl\{e^{-i\omega (t-t')}\left[\overrightarrow{R}^{(1)}_{\omega l}(r_i)\overrightarrow{R}^{(1 *)}_{\omega l}(r_j) + \overleftarrow{R}^{(1)}_{\omega l}(r_i)\overleftarrow{R}^{(1 *)}_{\omega l}(r_j)\right]
\nn\\
&+& e^{i\omega (t-t')}\left[\overrightarrow{R}^{(1)}_{\omega l}(r_j)\overrightarrow{R}^{(1 *)}_{\omega l}(r_i) + \overleftarrow{R}^{(1)}_{\omega l}(r_j)\overleftarrow{R}^{(1 *)}_{\omega l}(r_i)\right]\Biggr\},
\label{hada-rest-boul}
\eea
$i, j =1, 2$, where we have employed the addition theorem for the spherical harmonics~\cite{hilbert}
$$
\frac{4\pi}{2l + 1}\sum_{m = - l}^{l}Y_{lm}(\theta,\phi)Y^{*}_{lm}(\theta',\phi') = P_{l}(\hat{r} \cdot \hat{r}')
$$
where $\hat{r}$ and $\hat{r}'$ are two unit vectors with spherical coordinates $(\theta, \phi)$ and $(\theta', \phi')$, respectively, and $P_{l}$ is the Legendre polynomial of degree $l$~\cite{abram}. On the other hand, the Pauli-Jordan functions are given by
\bea
\Delta^{B}_{rr}(x_i(t),x_j(t')) &=& \frac{1}{16\pi^2}\sum_{l = 1}^{\infty}(2l +1)P_{l}(\hat{r}_i \cdot \hat{r}_j)
\int_{0}^{\infty}\,d\omega\,\omega\,\Biggl\{e^{-i\omega (t-t')}\left[\overrightarrow{R}^{(1)}_{\omega l}(r_i)\overrightarrow{R}^{(1 *)}_{\omega l}(r_j) + \overleftarrow{R}^{(1)}_{\omega l}(r_i)\overleftarrow{R}^{(1 *)}_{\omega l}(r_j)\right]
\nn\\
&-& e^{i\omega (t-t')}\left[\overrightarrow{R}^{(1)}_{\omega l}(r_j)\overrightarrow{R}^{(1 *)}_{\omega l}(r_i) + \overleftarrow{R}^{(1)}_{\omega l}(r_j)\overleftarrow{R}^{(1 *)}_{\omega l}(r_i)\right]\Biggr\}.
\label{pauli-rest-boul}
\eea
The contributions~(\ref{vfha3}) and~(\ref{rrha3}) to the rate of variation of atomic energy can be evaluated by inserting in such expressions the statistical functions of the two-atom system, given by equations~(\ref{susa1}) and~(\ref{cora1}), and the electromagnetic-field statistical functions given by~(\ref{hada-rest-boul}) and~(\ref{pauli-rest-boul}). Initially let us present the contributions coming from the vacuum fluctuations. Performing a simple change of variable $u = t - t'$, these can be expressed as, with $\Delta t = t - t_0$:
\bea
\Biggl\langle \frac{d H_A}{dt} \Biggr\rangle_{VF} &=& -\frac{1}{32\pi^2}\sum_{\nu'}\sum_{k,j = 1}^{2}\,
{\cal U}^{rr}_{kj}(\nu ,\nu')\,
\sqrt{g_{00}(r_k)g_{00}(r_j)}\,\exp\left[i\left(\widetilde{\Delta \nu}_k - \widetilde{\Delta \nu}_j\right)t\right]\,\Delta\nu
\nn\\
&\times&\int_{0}^{\infty}\,d\omega\,\omega
\,\Biggl\{C_{B}(\omega, r_k, r_j) \int_{0}^{\Delta t}du \left[e^{i\left(\widetilde{\Delta \nu}_j - \omega \right) u} 
+ e^{-i\left(\widetilde{\Delta \nu}_k  - \omega\right) u}\right] 
\nn\\
&+& C_{B}(\omega, r_j, r_k) \int_{0}^{\Delta t}du \,\left[e^{i\left(\widetilde{\Delta \nu}_j + \omega \right) u} 
+ e^{-i\left(\widetilde{\Delta \nu}_k  + \omega\right) u}\right]\Biggr\},
\eea
where $\widetilde{\Delta \nu}_i = \sqrt{g_{00}(r_i)}\,\Delta\nu$ (this comes from the usual gravitational redshift effect) and the generalized atomic transition dipole moment ${\cal U}_{ab}^{ij}(\nu, \nu')$ is given by equation~(\ref{mel}). Also we have defined
\beq
C_{B}(\omega, r, r') = \overrightarrow{C}_{B}(\omega, r, r') + \overleftarrow{C}_{B}(\omega, r, r'),
\eeq
with 
\bea
\overrightarrow{C}_{B}(\omega, r, r') &=& \sum_{l = 1}^{\infty} (2l+1)P_{l}(\hat{r} \cdot \hat{r}')\overrightarrow{R}^{(1)}_{\omega l}(r)\overrightarrow{R}^{(1 *)}_{\omega l}(r')
\nn\\
\overleftarrow{C}_{B}(\omega, r, r') &=& \sum_{l = 1}^{\infty} (2l+1)P_{l}(\hat{r} \cdot \hat{r}')\overleftarrow{R}^{(1)}_{\omega l}(r)\overleftarrow{R}^{(1 *)}_{\omega l}(r').
\eea
The above integrals can be easily solved and the result is
\bea
&&\Biggl\langle \frac{d H_A}{dt} \Biggr\rangle_{VF} = -\frac{1}{16\pi^2}\sum_{\nu'}\sum_{k,j = 1}^{2}\,
{\cal U}^{rr}_{kj}(\nu ,\nu')\,\sqrt{g_{00}(r_k)g_{00}(r_j)}\,\exp\left[i\left(\widetilde{\Delta \nu}_k - \widetilde{\Delta \nu}_j\right)t\right]\,\Delta\nu
\nn\\
&\times&\int_{0}^{\infty}\,d\omega\,\omega\,\Biggl\{C_{B}(\omega, r_k, r_j) \biggl[\frac{e^{i \Delta t/2 \left(\widetilde{\Delta \nu}_j -\omega\right)}\sin\left(\Delta t/2 \left(\widetilde{\Delta \nu}_j -\omega\right)\right)}{\widetilde{\Delta \nu}_j-\omega}
+\frac{e^{-i \Delta t/2 \left(\widetilde{\Delta \nu}_k -\omega\right)}\sin\left(\Delta t/2 \left(\widetilde{\Delta \nu}_k -\omega\right)\right)}{\widetilde{\Delta \nu}_k -\omega}\biggr] 
\nn\\
&+& C_{B}(\omega, r_j, r_k) \biggl[\frac{e^{i \Delta t/2 \left(\widetilde{\Delta \nu}_j +\omega\right)}\left(\sin\left(\Delta t/2 \left(\widetilde{\Delta \nu}_j +\omega\right)\right)\right)}{\widetilde{\Delta \nu}_j+\omega}
+ \frac{e^{-i \Delta t/2 \left(\widetilde{\Delta \nu}_k +\omega\right)}\sin\left(\Delta t/2 \left(\widetilde{\Delta \nu}_k +\omega\right)\right)}{\widetilde{\Delta \nu}_k+\omega}\biggr] \Biggr\}.
\nn\\
\label{vacuum-b}
\eea
For sufficiently large $\Delta t$, $\sin(\Delta t\,x)/x \to \pi \delta(x)$ and the integral over $\omega$ can be explicitly solved. In this limit, it becomes clear to note that vacuum fluctuations tend to excite ($\widetilde{\Delta \nu}_i < 0 \Rightarrow \langle d H_A/dt \rangle_{VF} > 0$) as well as deexcite ($\widetilde{\Delta \nu}_i > 0 \Rightarrow \langle d H_A/dt \rangle_{VF} < 0$) the atomic system. In the present context, this means that the atoms disentangle and can also entangle in a finite observation time due to vacuum fluctuations of the electromagnetic field. The creation of entanglement due to vacuum fluctuations persists even at late times, as this result plainly shows.

Now let us present the radiation-reaction contributions. Performing similar calculations as above one gets:
\bea
&&\Biggl\langle \frac{d H_A}{dt} \Biggr\rangle_{RR} = -\frac{1}{16\pi^2}\sum_{\nu'}\sum_{k,j = 1}^{2}\,
{\cal U}^{rr}_{kj}(\nu ,\nu')\,\sqrt{g_{00}(r_k)g_{00}(r_j)}\,\exp\left[i\left(\widetilde{\Delta \nu}_k - \widetilde{\Delta \nu}_j\right)t\right]\,\Delta\nu
\nn\\
&\times&\int_{0}^{\infty}\,d\omega\,\omega\,\Biggl\{C_{B}(\omega, r_k, r_j) \biggl[\frac{e^{i \Delta t/2 \left(\widetilde{\Delta \nu}_j -\omega\right)}\sin\left(\Delta t/2 \left(\widetilde{\Delta \nu}_j -\omega\right)\right)}{\widetilde{\Delta \nu}_j-\omega}
+\frac{e^{-i \Delta t/2 \left(\widetilde{\Delta \nu}_k -\omega\right)}\sin\left(\Delta t/2 \left(\widetilde{\Delta \nu}_k -\omega\right)\right)}{\widetilde{\Delta \nu}_k -\omega}\biggr] 
\nn\\
&-& C_{B}(\omega, r_j, r_k) \biggl[\frac{e^{i \Delta t/2 \left(\widetilde{\Delta \nu}_j +\omega\right)}\left(\sin\left(\Delta t/2 \left(\widetilde{\Delta \nu}_j +\omega\right)\right)\right)}{\widetilde{\Delta \nu}_j+\omega}
+ \frac{e^{-i \Delta t/2 \left(\widetilde{\Delta \nu}_k +\omega\right)}\sin\left(\Delta t/2 \left(\widetilde{\Delta \nu}_k +\omega\right)\right)}{\widetilde{\Delta \nu}_k+\omega}\biggr] \Biggr\}.
\nn\\
\label{radiation-b}
\eea
Observe that the effect of radiation reaction always leads to a loss of atomic energy $\langle d H_A/dt \rangle_{VF} < 0$ independent of how the atomic system were initially prepared. In other words, with respect to absorption processes the radiation reaction do not contribute to the generation of entanglement between the atoms; in this case, it always tend to disentangle an entangled state via spontaneous emission processes. This is reminiscent from the fact that classical noise coupled to an entangled quantum two-level system will generally lead to decoherence and disentanglement processes. 

For completeness, let us present the total rate of change of the atomic energy. This is obtained by adding the contributions of vacuum fluctuations and radiation reaction. One gets
\bea
&&\Biggl\langle \frac{d H_A}{dt} \Biggr\rangle_{tot} = -\frac{1}{8\pi^2}\sum_{\nu'}\sum_{k,j = 1}^{2}\,
{\cal U}^{rr}_{kj}(\nu ,\nu')\,\sqrt{g_{00}(r_k)g_{00}(r_j)}\,\exp\left[i\left(\widetilde{\Delta \nu}_k - \widetilde{\Delta \nu}_j\right)t\right]\,\Delta \nu
\nn\\
&\times&\int_{0}^{\infty}\,d\omega\,\omega\,C_{B}(\omega, r_k, r_j) \biggl[\frac{e^{i \Delta t/2 \left(\widetilde{\Delta \nu}_j -\omega\right)}\sin\left(\Delta t/2 \left(\widetilde{\Delta \nu}_j -\omega\right)\right)}{\widetilde{\Delta \nu}_j-\omega}
+\frac{e^{-i \Delta t/2 \left(\widetilde{\Delta \nu}_k -\omega\right)}\sin\left(\Delta t/2 \left(\widetilde{\Delta \nu}_k -\omega\right)\right)}{\widetilde{\Delta \nu}_k -\omega}\biggr].
\nn\\
\label{total-b}
\eea
\end{widetext}
The result clearly shows that it is possible to generate entanglement between the atoms via absorption process for a finite observation time in the case of a quantum electromagnetic field in the Boulware vacuum state. However, once entanglement is created it lasts only a finite duration, and always disappears at late observation times $\Delta t$ (see the remark above concerning the $\omega$-integral for large $\Delta t$). This is a similar result as the one found in reference~\cite{LinHu2010}, yet in a entirely different scenario. Typically this entangled state lasts a duration of order $\sim 1/(\widetilde{\Delta \nu}_i -\omega)$ which corresponds roughly to the greatest width of the peaks of the functions $\sin x/x$ in the above integrand. In other words, one gets a finite result only for $\Delta\nu > 0$ for asymptotic $\Delta t$: the balance between vacuum fluctuations and radiation reaction prevents the atoms from getting entangled via an absorption process. In addition, for a finite observation time notice that the situation in which $\Delta t < |\Delta{\bf x}|$, $|\Delta{\bf x}|$ being the distance between the atoms, is allowed. This does not bring any controversial issues regarding causality since it is widely known that entangled quantum states produce nonlocal correlations~\cite{bus}. Furthermore, note from equation~(\ref{total-b}) with large $\Delta t$ that as the atoms approach each other, one gets $\langle d H_A/dt\rangle \to 0$ for atoms initially prepared in the entangled state $|\Omega^{-}\rangle$ with $c_1 = c_2 = 1/\sqrt{2}$, which means that such a state is stable with respect to radiative processes. Thus we recover the well known result which states that atoms confined into a region much smaller than the optical wavelength, the antisymmetric entangled state $|\Omega^{-}\rangle$ with $c_1 = c_2 = 1/\sqrt{2}$ can be regarded as a decoherence-free state~\cite{rep}.

Let us briefly discuss the rate of change of the atomic energy for the asymptotic regions of interest. All the relevant calculations are presented in detail in Appendix~\ref{B}. For simplicity, assume a large enough $\Delta t$ so that one could approximate $\sin x/x$ as delta functions. First let us consider the asymptotic region $r_1, r_2 \to \infty$. From the results derived in Appendix~\ref{B} one gets:
\beq
C_{B}(\omega, r_i, r_j) \approx H_{+}(\omega, r_i, r_j)+ F(\omega, {\bf x}_i,{\bf x}_j)
\eeq
where the function $F(\omega, {\bf x},{\bf x}')$ is given by equation~(\ref{f}) and we have defined
\bea
H_{\pm}(\omega, r, r') &=& \sum_{l=1}^{\infty}l(l+1)(2l + 1)\,P_{l}(\hat{r} \cdot \hat{r}')|{\cal B}_{l}(\omega)|^2
\nn\\
&\times&\,\frac{e^{\pm i\omega (r_{*} - r'_{*})}}{\omega^2 r^2 r'^2},
\label{H}
\eea
where ${\cal B}_{l}(\omega)$ is the usual transmission coefficient defined through equations~(\ref{asymp}) and $r_{*} = r + 2M\ln(r/2M - 1)$ is the Regge-Wheeler tortoise coordinate. For estimation of the sum on equation~(\ref{H}) one can study the gravitational capture cross-section of test particles whose trajectories terminates in the black hole~\cite{frolov}. One finds that, if the impact parameter $b$ of a ultra-relativistic particle coming in from infinity is less than the critical value $\sqrt{27} M$, such a particle gets captured by the black hole. Employing the relation $l = \omega b$, one rewrites the capture condition as $l < \sqrt{27} M\omega$. Hence assuming that all modes obeying such a relation are absorbed by the black hole, one can suitably approximate the transmission coefficient by $|{\cal B}_{l}(\omega)|^2 \sim \theta(\sqrt{27} M\omega - l)$, where $\theta(z)$ is the usual Heaviside step function. This is sometimes called DeWitt approximation~\cite{dewitt} but it is essentially a geometrical optics approximation for all wavelengths. Hence one gets:
\begin{widetext}
\bea
H_{\pm}(\omega, r, r') &\approx& \sum_{l=1}^{\infty}\frac{l(l+1)(2l + 1)\,P_{l}(\hat{r} \cdot \hat{r}')\theta(\sqrt{27} M\omega - l) e^{\pm i\omega (r_{*} - r'_{*})}}{\omega^2 r^2 r'^2} 
\approx \frac{2\,e^{\pm i\omega (r_{*} - r'_{*})}}{\omega^2 r^2 r'^2}\int_{0}^{\sqrt{27} M\omega}dl\,l^3 J_0(l\gamma),
\label{H1}
\eea
\end{widetext}
where $\cos\gamma  = {\bf \hat{r}} \cdot {\bf \hat{r}}'$ and we have employed the asymptotic result: 
$P_{\nu}[\cos (x/\nu)] \approx J_{0}(x) + {\cal O}(\nu^{-1})$, with $J_{\mu}(x)$ being the Bessel function of the first kind~\cite{abram}. We distinguish two separate cases. For ${\bf r} = {\bf r}'$ one gets
\bea
H_{\pm}(\omega, r, r) &\approx& \frac{729\,M^4\omega^2}{2\, r^4},
\label{H2a}
\eea
and for ${\bf r} \neq {\bf r}'$ one gets
\bea
H_{\pm}(\omega, r, r') &\approx& \frac{54 M^2\,e^{\pm i\omega (r_{*} - r'_{*})}}{\gamma ^2 r^2 r'^2}
\nn\\
&\times&\,\left[2\,J_2(z(\omega)) - z(\omega)\, J_3(z(\omega))\right],
\label{H2b}
\eea
where we have defined the function $z(\omega) = \sqrt{27} M \gamma \omega$ and the following integral was used~\cite{prud}
$$
\int_{0}^{x} dy\,y^3 J_0(y) = x^2 [2 J_2(x) - x J_3(x)].
$$
It is easy to see that, at infinity, $H_{\pm}$ gives vanishingly small contributions, so that the rate of change reduces essentially to that of inertial atoms in the Minkowski vacuum in flat space-times with no boundaries. In this way the results of reference~\cite{ng1} are reproduced. Note also that $F({\omega}, {\bf x},{\bf x}') \sim 0$ for large distance between the atoms, but it is finite for $r_1 = r_2 \to \infty$, see equation~(\ref{asymp-inf4}). This means that for large asymptotic separation between the atoms the cross correlations arising in $\langle d H_A/dt \rangle$ vanish and one is left with terms corresponding to isolated atoms. 

The other important region is when $r_1, r_2 \to 2M$. One has that
\bea
C_{B}(\omega, r_i, r_j) &\approx& H_{-}(\omega, r_i, r_j)+ \frac{16 a^2(r_i) \sinh(\pi \xi(\omega))}{g_{00}(r_i)\,\pi \xi(\omega)}
\nn\\
&\times&\,\biggl[A_{i\xi(\omega)}(\gamma, g_{00}(r_j), g_{00}(r_i)) 
\nn\\
&+&\, A_{-i\xi(\omega)}(\gamma, g_{00}(r_j), g_{00}(r_i))\biggr],
\eea
where $a(r) = M/(r^2 \sqrt{g_{00}(r)}\,)$ is the proper acceleration of the static atom at $r$,  $\xi(\omega) = 4M\omega$ and $A_{i\xi}(\gamma, r, r') = A_{i\xi}(\gamma, g_{00}(r), g_{00}(r'))$ is properly defined in the Appendix~\ref{B}. For finite $\Delta t$, as in the previous case one gets a finite result regardless of the sign of $\Delta\nu$. In addition, note that $g_{00}(r)$ vanishes as the event horizon is approached, thence the rate of change of the atomic energy diverges.  

As a last analysis concerning the Boulware vacuum we take, say, $r_2 \to 2M$ whereas $r_1$ is kept arbitrary. One gets:
\bea
&&C_{B}(\omega, r_1, r_2) \approx \sum_{l=1}^{\infty}\frac{l(l+1) (2l + 1)\,P_{l}(\hat{r}_1 \cdot \hat{r}_2)}{(r_2)^2 (r_1)^2 \omega^2} 
\nn\\
&&\times\,\biggl[\overleftarrow{{\cal R}}^{(1)}_{\omega l}(r_1){\cal B}^{*}_{l}(\omega) e^{i\omega  r_{2*}} 
\nn\\
&&+\, \frac{2e^{-i\omega/2\kappa}}{\Gamma(i\omega/\kappa)} \overrightarrow{{\cal R}}^{(1)}_{\omega l}(r_1) e^{i\omega \ln l/\kappa} K_{i\omega/\kappa}\left(2l\sqrt{g_{00}(r_2)}\right)\biggr]
\eea
where $\kappa = 1/4M$, and a similar result for ${\cal C}_{B}(\omega, r_2, r_1)$, see equation~(\ref{asymp-2m5}). From the results found in Appendix~\ref{B}, as $r_1 \to \infty$ the cross terms vanish and again we are left only with terms corresponding to isolated atoms. On the other hand, as $r_1$ approaches $2M$, $\langle d H_A/dt \rangle$ diverges and we recover the previous results discussed. Observe the general result: as the atoms approach $r = 2M$ the rate of variation of atomic energy grows rapidly and violently. For large $\Delta t$ this implies a vastly fast degradation of entanglement between the atoms initially prepared in one of the entangled states $|\Omega^{\pm}\rangle$.

\subsection{The Hartle-Hawking vacuum}

The Hartle-Hawking vacuum state is not empty at infinity, corresponding to a thermal distribution of quanta at the black-hole temperature. In other words, the Hartle-Hawking vacuum describes the physical situation in which the black hole is in equilibrium with an infinite sea of black-body radiation, such as would be observed by constraining the black hole to the interior of a perfectly reflecting cavity. The renormalized expectation value of the stress tensor is well behaved in a freely falling frame on the horizon~\cite{sciama}.

\begin{widetext}

We now proceed to calculate the rate of variation of atomic energy in the Hartle-Hawking vacuum state. From the results derived in the Appendix~\ref{A}, one may compute all the relevant correlation functions of the electric field which appears in equations~(\ref{vfha3}) and~(\ref{rrha3}). The associated Hadamard's elementary functions are given by
\bea
D^{H}_{rr}(x_i(t),x_j(t')) &=& \frac{1}{16\pi^2}\sum_{l = 1}^{\infty}(2l +1)P_{l}(\hat{r}_i \cdot \hat{r}_j)
\int_{-\infty}^{\infty}\,d\omega\,\omega\,\Biggl\{e^{-i\omega (t-t')} 
\left[\frac{\overrightarrow{R}^{(1)}_{\omega l}(r_i)\overrightarrow{R}^{(1 *)}_{\omega l}(r_j)}{1 - e^{-2\pi\omega/\kappa}} + \frac{\overleftarrow{R}^{(1)}_{\omega l}(r_i)\overleftarrow{R}^{(1 *)}_{\omega l}(r_j)}{e^{2\pi\omega/\kappa} - 1}\right]
\nn\\
&& + \,e^{i\omega (t-t')} 
\left[\frac{\overrightarrow{R}^{(1)}_{\omega l}(r_j)\overrightarrow{R}^{(1 *)}_{\omega l}(r_i)}{1 - e^{-2\pi\omega/\kappa}} + \frac{\overleftarrow{R}^{(1)}_{\omega l}(r_j)\overleftarrow{R}^{(1 *)}_{\omega l}(r_i)}{e^{2\pi\omega/\kappa} - 1}\right]\Biggr\},
\label{hada-rest-haw}
\eea
$i, j =1, 2$, where we have employed the addition theorem for the spherical harmonics quoted above. On the other hand, the Pauli-Jordan functions are given by
\bea
\Delta^{H}_{rr}(x_1i(t),x_j(t')) &=& \frac{1}{16\pi^2}\sum_{l = 1}^{\infty}(2l +1)P_{l}(\hat{r}_i \cdot \hat{r}_j)
\int_{-\infty}^{\infty}\,d\omega\,\omega\,\Biggl\{e^{-i\omega (t-t')} 
\left[\frac{\overrightarrow{R}^{(1)}_{\omega l}(r_i)\overrightarrow{R}^{(1 *)}_{\omega l}(r_j)}{1 - e^{-2\pi\omega/\kappa}} - \frac{\overleftarrow{R}^{(1)}_{\omega l}(r_i)\overleftarrow{R}^{(1 *)}_{\omega l}(r_j)}{e^{2\pi\omega/\kappa} - 1}\right]
\nn\\
&& - \,e^{i\omega (t-t')} 
\left[\frac{\overrightarrow{R}^{(1)}_{\omega l}(r_j)\overrightarrow{R}^{(1 *)}_{\omega l}(r_i)}{1 - e^{-2\pi\omega/\kappa}} - \frac{\overleftarrow{R}^{(1)}_{\omega l}(r_j)\overleftarrow{R}^{(1 *)}_{\omega l}(r_i)}{e^{2\pi\omega/\kappa} - 1}\right]\Biggr\}.
\label{pauli-rest-haw}
\eea
Now such expressions as well as the statistical functions of the two-atom system, given by equations~(\ref{susa1}) and~(\ref{cora1}), should be inserted in equations~(\ref{vfha3}) and~(\ref{rrha3}). As above we begin with the contributions coming from the vacuum fluctuations. Performing a simple change of variable $u = t - t'$, these can be expressed as, with $\Delta t = t - t_0$:
\bea
\Biggl\langle \frac{d H_A}{dt} \Biggr\rangle_{VF} &=& -\frac{1}{32\pi^2}\sum_{\nu'}\sum_{k,j = 1}^{2}\,
{\cal U}^{rr}_{kj}(\nu ,\nu')\,\sqrt{g_{00}(r_k)g_{00}(r_j)}\,\exp\left[i\left(\widetilde{\Delta \nu}_k - \widetilde{\Delta \nu}_j\right)t\right]\,\Delta\nu
\nn\\
&\times&\int_{-\infty}^{\infty}\,d\omega\,\omega\,\Biggl\{C^{+}_{H}(\omega, r_k, r_j) \int_{0}^{\Delta t}du \left[e^{i\left(\widetilde{\Delta \nu}_j - \omega \right) u} 
+ e^{-i\left(\widetilde{\Delta \nu}_k  - \omega\right) u}\right] 
\nn\\
&+& C^{+}_{H}(\omega, r_j, r_k) \int_{0}^{\Delta t}du \,\left[e^{i\left(\widetilde{\Delta \nu}_j + \omega \right) u} 
+ e^{-i\left(\widetilde{\Delta \nu}_k  + \omega\right) u}\right]\Biggr\},
\eea
where we have defined
\beq
C^{\pm}_{H}(\omega, r, r') = \overrightarrow{C}_{H}(\omega, r, r') \pm \overleftarrow{C}_{H}(\omega, r, r'),
\eeq
with 
\bea
\overrightarrow{C}_{H}(\omega, r, r') &=& \sum_{l = 1}^{\infty} (2l+1)P_{l}(\hat{r} \cdot \hat{r}')\,\overrightarrow{R}^{(1)}_{\omega l}(r)\overrightarrow{R}^{(1 *)}_{\omega l}(r')\left(1 +\frac{1}{e^{2\pi\omega/\kappa} - 1}\right)
\nn\\
\overleftarrow{C}_{H}(\omega, r, r') &=& \sum_{l = 1}^{\infty} (2l+1)P_{l}(\hat{r} \cdot \hat{r}')\frac{\overleftarrow{R}^{(1)}_{\omega l}(r)\overleftarrow{R}^{(1 *)}_{\omega l}(r')}{e^{2\pi\omega/\kappa} - 1}.
\eea
Solving the above integrals leads us to the following result:
\bea
&&\Biggl\langle \frac{d H_A}{dt} \Biggr\rangle_{VF} = -\frac{1}{16\pi^2}\sum_{\nu'}\sum_{k,j = 1}^{2}\,
{\cal U}^{rr}_{kj}(\nu ,\nu')\,\sqrt{g_{00}(r_k)g_{00}(r_j)}\,\exp\left[i\left(\widetilde{\Delta \nu}_k - \widetilde{\Delta \nu}_j\right)t\right]\,\Delta\nu
\nn\\
&\times&\int_{-\infty}^{\infty}\,d\omega\,\omega\,\Biggl\{C^{+}_{H}(\omega, r_k, r_j) \biggl[\frac{e^{i \Delta t/2 \left(\widetilde{\Delta \nu}_j -\omega\right)}\sin\left(\Delta t/2 \left(\widetilde{\Delta \nu}_j -\omega\right)\right)}{\widetilde{\Delta \nu}_j-\omega}
+\frac{e^{-i \Delta t/2 \left(\widetilde{\Delta \nu}_k -\omega\right)}\sin\left(\Delta t/2 \left(\widetilde{\Delta \nu}_k -\omega\right)\right)}{\widetilde{\Delta \nu}_k -\omega}\biggr] 
\nn\\
&+& C^{+}_{H}(\omega, r_j, r_k) \biggl[\frac{e^{i \Delta t/2 \left(\widetilde{\Delta \nu}_j +\omega\right)}\left(\sin\left(\Delta t/2 \left(\widetilde{\Delta \nu}_j +\omega\right)\right)\right)}{\widetilde{\Delta \nu}_j+\omega}
+ \frac{e^{-i \Delta t/2 \left(\widetilde{\Delta \nu}_k +\omega\right)}\sin\left(\Delta t/2 \left(\widetilde{\Delta \nu}_k +\omega\right)\right)}{\widetilde{\Delta \nu}_k+\omega}\biggr] \Biggr\}.
\nn\\
\label{vacuum-h}
\eea
Observe the appearance of the thermal terms coming from the function $C^{+}_{H}$. This is most readily seen by letting $\Delta t$ approaches asymptotic values, which leads to delta functions in the above expressions and again the integral over $\omega$ can be explicitly solved. As for the Boulware vacuum state, vacuum fluctuations tend to generate entanglement between atoms initially prepared in the ground state which is sustained over late periods of observational time. Similarly, vacuum fluctuations may destroy initially entangled atoms. Both processes are ensured with equal magnitude and are heightened by the thermal terms compared
to the Boulware case.

Now let us present the radiation-reaction contributions. The calculations follow similar steps as above and the result is
\bea
&&\Biggl\langle \frac{d H_A}{dt} \Biggr\rangle_{RR} = -\frac{1}{16\pi^2}\sum_{\nu'}\sum_{k,j = 1}^{2}\,
{\cal U}^{rr}_{kj}(\nu ,\nu')\,\sqrt{g_{00}(r_k)g_{00}(r_j)}\,\exp\left[i\left(\widetilde{\Delta \nu}_k - \widetilde{\Delta \nu}_j\right)t\right]\,\Delta\nu
\nn\\
&\times&\int_{-\infty}^{\infty}\,d\omega\,\omega\,\Biggl\{C^{-}_{H}(\omega, r_k, r_j) \biggl[\frac{e^{i \Delta t/2 \left(\widetilde{\Delta \nu}_j -\omega\right)}\sin\left(\Delta t/2 \left(\widetilde{\Delta \nu}_j -\omega\right)\right)}{\widetilde{\Delta \nu}_j-\omega}
+\frac{e^{-i \Delta t/2 \left(\widetilde{\Delta \nu}_k -\omega\right)}\sin\left(\Delta t/2 \left(\widetilde{\Delta \nu}_k -\omega\right)\right)}{\widetilde{\Delta \nu}_k -\omega}\biggr] 
\nn\\
&-& C^{-}_{H}(\omega, r_j, r_k) \biggl[\frac{e^{i \Delta t/2 \left(\widetilde{\Delta \nu}_j +\omega\right)}\left(\sin\left(\Delta t/2 \left(\widetilde{\Delta \nu}_j +\omega\right)\right)\right)}{\widetilde{\Delta \nu}_j+\omega}
+ \frac{e^{-i \Delta t/2 \left(\widetilde{\Delta \nu}_k +\omega\right)}\sin\left(\Delta t/2 \left(\widetilde{\Delta \nu}_k +\omega\right)\right)}{\widetilde{\Delta \nu}_k+\omega}\biggr] \Biggr\}.
\nn\\
\label{radiation-h}
\eea
Note that the contribution from radiation reaction is also altered by the appearance of the thermal terms encoded in $C^{-}_{H}$. This is in sharp contrast with the situation of uniformly accelerated atoms coupled with a quantum field prepared in Minkowski vacuum~\cite{ng2} (for a related result, see Ref.~\cite{china}). Nevertheless, as in the Boulware case, the radiation reaction do not contribute to the generation of entanglement between the atoms through absorption process, leading always to disentanglement via spontaneous emission processes.

The total rate of change of the atomic energy is obtained by adding the contributions of vacuum fluctuations and radiation reaction. One gets, for sufficiently large $\Delta t$:
\bea
&&\Biggl\langle \frac{d H_A}{dt} \Biggr\rangle_{tot} = -\frac{1}{8\pi}\sum_{\nu'}\sum_{k,j = 1}^{2}\,
{\cal U}^{rr}_{kj}(\nu ,\nu')\,\sqrt{g_{00}(r_k)g_{00}(r_j)}\,\exp\left[i\left(\widetilde{\Delta \nu}_k - \widetilde{\Delta \nu}_j\right)t\right]\,\Delta\nu
\nn\\
&\times&\,\Biggl[\widetilde{\Delta \nu}_j\,\overrightarrow{C}_{H}(\widetilde{\Delta \nu}_j, r_k, r_j)
+ \widetilde{\Delta \nu}_k\,\overrightarrow{C}_{H}(\widetilde{\Delta \nu}_k, r_k, r_j) 
- \widetilde{\Delta \nu}_j\,\overleftarrow{C}_{H}(-\widetilde{\Delta \nu}_j, r_j, r_k) - 
\widetilde{\Delta \nu}_k\,\overleftarrow{C}_{H}(-\widetilde{\Delta \nu}_k, r_j, r_k) \Biggr].
\label{total-h}
\eea
Observe from equation~(\ref{total-h}) that as the atoms approach each other, the entangled state $|\Omega^{-}\rangle$ with $c_1 = c_2 = 1/\sqrt{2}$ is again stable with respect to radiative processes. 
\end{widetext}

The balance between vacuum fluctuations and radiation reaction which existed in the Boulware vacuum is disturbed and entanglement can be created via absorption process in the exterior region of the black hole even for large asymptotic $\Delta t$. In other words, both possibilities $\Delta\nu > 0$ and $\Delta\nu < 0$ are allowed. The Planckian factors which appears in equation~(\ref{total-h}) through the functions $\overrightarrow{C}_{H}$ and $\overleftarrow{C}_{H}$ uncovers the thermal nature of the Hartle-Hawking vacuum. For large enough $\Delta t$, one easily sees that the temperature of the thermal radiation is given by the Hawking temperature:
\beq
T^{H}_{i} = \frac{\kappa}{2\pi \sqrt{g_{00}(r_i)}} = \frac{1}{8\pi M\,\sqrt{1-2M/r_i}},
\label{t-haw}
\eeq
which is just the Tolman relation which gives the temperature felt by a local observer at the fixed position $r_i$~\cite{tolman}. The emergence of two different temperatures is a feature of the atoms being at different fixed positions. Even though we find different temperatures, the thermal equilibrium is warranted by invoking the Tolman relation $(g_{00}(x))^{1/2}\,T(x) = \textrm{const}$. 

Further inquires must be answered by inspecting the result in the asymptotic regions. Consider a large enough $\Delta t$ as above. For the atoms fixed at spatial infinity, i.e. $r_1, r_2 \to \infty$, the results discussed in Appendix~\ref{B} reveals that:
\bea
C^{\pm}_{H}(\omega, r_i, r_j) &\approx& H_{+}(\omega, r_i, r_j)\left(1 +\frac{1}{e^{2\pi\omega/\kappa} - 1}\right) 
\nn\\
&\pm&\,  \frac{F(\omega, {\bf x}_i,{\bf x}_j)}{e^{2\pi\omega/\kappa} - 1},
\eea
where the function $F(\omega, {\bf x},{\bf x}')$ is given by equation~(\ref{f}) and $H_{\pm}(\omega, r, r')$ is given by expression~(\ref{H}) (or equations~(\ref{H2a}),~(\ref{H2b}) within the geometrical optics approximation discussed above). At infinity and for large distance between the atoms, $H_{\pm}(\omega, r, r')$ and $F(\omega, {\bf x},{\bf x}')$ give vanishingly small contributions and we are left with a summation of terms related with isolated atoms each one following its own world line. Recalling the thermalization theorem~\cite{sciama,fulling,davies}, one is led to the conclusion that in the situation with sufficiently high relative asymptotic distance we have two atoms coupled individually to two spatially separated cavities at different Hawking temperatures in a flat space-time. Hence our results indicates a close resemblance with the outcomes of Ref.~\cite{eberly}.

At the vicinity of the event horizon, i.e., $r_1, r_2 \to 2M$, one has that
\begin{widetext}
\bea
C^{\pm}_{H}(\omega, r_i, r_j) &\approx& \frac{16 a^2(r_i) \sinh(\pi \xi(\omega))}{g_{00}(r_i)\,\pi \xi(\omega)}
\left[A_{i\xi(\omega)}(\gamma, g_{00}(r_j), g_{00}(r_i)) + A_{-i\xi(\omega)}(\gamma, g_{00}(r_j), g_{00}(r_i))\right]\left(1 +\frac{1}{e^{2\pi\omega/\kappa} - 1}\right) 
\nn\\
&\pm&\, \frac{H_{-}(\omega, r_i, r_j)}{e^{2\pi\omega/\kappa} - 1},
\eea
\end{widetext}
which clearly shows a divergent result for $a(r) \to \infty$. We also observe two kinds of contributions, namely one related with the outgoing thermal radiation from the event horizon, and the other one associated with the thermal term multiplied by $H_{-}$, which can be interpreted as a consequence of existence of incoming thermal radiation from infinity. It is precisely this thermal nature that enables the atoms to get entangled even if they were initially prepared in the separable ground state.

Finally consider that $r_2 \to 2M$ whereas $r_1$ is kept arbitrary. One gets:
\begin{widetext}
\bea
C^{\pm}_{H}(\omega, r_1, r_2) &\approx& \sum_{l=1}^{\infty}\frac{l(l+1) (2l + 1)\,P_{l}(\hat{r}_1 \cdot \hat{r}_2)}{(r_2)^2 (r_1)^2 \omega^2} 
\biggl[\frac{2e^{-i\omega/2\kappa}}{\Gamma(i\omega/\kappa)} \overrightarrow{{\cal R}}^{(1)}_{\omega l}(r_1) e^{i\omega \ln l/\kappa} K_{i\omega/\kappa}\left(2l\sqrt{g_{00}(r_2)}\right)\left(1 +\frac{1}{e^{2\pi\omega/\kappa} - 1}\right) 
\nn\\
&\pm&\, \frac{\overleftarrow{{\cal R}}^{(1)}_{\omega l}(r_1){\cal B}^{*}_{l}(\omega) e^{i\omega  r_{*2}}}{e^{2\pi\omega/\kappa} - 1}\biggr].
\eea
\end{widetext}
and a similar result for ${\cal C}^{\pm}_{H}(\omega, r_2, r_1)$, see equation~(\ref{asymp-2m5}). From the results found in Appendix~\ref{B}, as $r_1 \to \infty$ the cross terms vanish and again we are left only with terms corresponding to isolated atoms. On the other hand, as $r_1$ approaches $2M$, $\langle d H_A/dt \rangle$ diverges and we recover the previous results discussed. Again we have obtained the general result aforementioned: as the atoms approach the event horizon the rate of variation of atomic energy grows rapidly and violently. For large $\Delta t$ this implies a greatly enhanced generation of entanglement between the atoms initially prepared in the ground state.

\subsection{The Unruh vacuum}

The Unruh vacuum state is the adequate choice of vacuum state which is most relevant to the gravitational collapse of a massive body. At spatial infinity this vacuum is tantamount to an outgoing flux of black-body radiation at the black-hole temperature. The renormalized expectation value of the stress tensor, in a freely falling frame, is well behaved on the future horizon but not on the past horizon~\cite{sciama}.

We now proceed to calculate the rate of variation of atomic energy in the Unruh vacuum state. From the results derived in the Appendix~\ref{A}, one may compute all the relevant correlation functions of the electric field which appears in equations~(\ref{vfha3}) and~(\ref{rrha3}). The associated Hadamard's elementary functions reads
\begin{widetext}
\bea
D^{U}_{rr}(x_i(t),x_j(t')) &=& \frac{1}{16\pi^2}\sum_{l = 1}^{\infty}(2l +1)P_{l}(\hat{r}_i \cdot \hat{r}_j)
\int_{-\infty}^{\infty}\,d\omega\,\omega\,\Biggl\{e^{-i\omega (t-t')}\left[\frac{\overrightarrow{R}^{(1)}_{\omega l}(r_i)\overrightarrow{R}^{(1 *)}_{\omega l}(r_j)}{1 - e^{-2\pi\omega/\kappa}} + \theta(\omega)\overleftarrow{R}^{(1)}_{\omega l}(r_i)\overleftarrow{R}^{(1 *)}_{\omega l}(r_j)\right]
\nn\\
&+& e^{i\omega (t-t')}\left[\frac{\overrightarrow{R}^{(1)}_{\omega l}(r_j)\overrightarrow{R}^{(1 *)}_{\omega l}(r_i)}{1 - e^{-2\pi\omega/\kappa}} + \theta(\omega)\overleftarrow{R}^{(1)}_{\omega l}(r_j)\overleftarrow{R}^{(1 *)}_{\omega l}(r_i)\right]\Biggr\},
\label{hada-rest-u}
\eea
$i, j =1, 2$, where use was made of the foregoing addition theorem for the spherical harmonics. In turn, the Pauli-Jordan functions are given by
\bea
\Delta^{U}_{rr}(x_i(t),x_j(t')) &=& \frac{1}{16\pi^2}\sum_{l = 1}^{\infty}(2l +1)P_{l}(\hat{r}_i \cdot \hat{r}_j)
\int_{-\infty}^{\infty}\,d\omega\,\omega\,\Biggl\{e^{-i\omega (t-t')}\left[\frac{\overrightarrow{R}^{(1)}_{\omega l}(r_i)\overrightarrow{R}^{(1 *)}_{\omega l}(r_j)}{1 - e^{-2\pi\omega/\kappa}} + \theta(\omega)\overleftarrow{R}^{(1)}_{\omega l}(r_i)\overleftarrow{R}^{(1 *)}_{\omega l}(r_j)\right]
\nn\\
&-& e^{i\omega (t-t')}\left[\frac{\overrightarrow{R}^{(1)}_{\omega l}(r_j)\overrightarrow{R}^{(1 *)}_{\omega l}(r_i)}{1 - e^{-2\pi\omega/\kappa}} + \theta(\omega)\overleftarrow{R}^{(1)}_{\omega l}(r_j)\overleftarrow{R}^{(1 *)}_{\omega l}(r_i)\right]\Biggr\}.
\label{pauli-rest-u}
\eea
The contributions~(\ref{vfha3}) and~(\ref{rrha3}) to the rate of variation of atomic energy can be evaluated by inserting in such expressions the statistical functions of the two-atom system, given by equations~(\ref{susa1}) and~(\ref{cora1}), and the electromagnetic-field statistical functions given by~(\ref{hada-rest-u}) and~(\ref{pauli-rest-u}). Initially let us focus on the contributions coming from the vacuum fluctuations. Performing a simple change of variable $u = t - t'$, these can be expressed as, with $\Delta t = t - t_0$:
\bea
\Biggl\langle \frac{d H_A}{dt} \Biggr\rangle_{VF} &=& -\frac{1}{32\pi^2}\sum_{\nu'}\sum_{k,j = 1}^{2}\,
{\cal U}^{rr}_{kj}(\nu ,\nu')\,\sqrt{g_{00}(r_k)g_{00}(r_j)}\,\exp\left[i\left(\widetilde{\Delta \nu}_k - \widetilde{\Delta \nu}_j\right)t\right]\,\Delta\nu
\nn\\
&\times&\int_{-\infty}^{\infty}\,d\omega\,\omega\,\Biggl\{C_{U}(\omega, r_k, r_j) \int_{0}^{\Delta t}du \left[e^{i\left(\widetilde{\Delta \nu}_j - \omega \right) u} 
+ e^{-i\left(\widetilde{\Delta \nu}_k  - \omega\right) u}\right] 
\nn\\
&+& C_{U}(\omega, r_j, r_k) \int_{0}^{\Delta t}du \,\left[e^{i\left(\widetilde{\Delta \nu}_j + \omega \right) u} 
+ e^{-i\left(\widetilde{\Delta \nu}_k  + \omega\right) u}\right]\Biggr\},
\eea
where we have defined
\beq
C_{U}(\omega, r, r') = \overrightarrow{C}_{U}(\omega, r, r') + \overleftarrow{C}_{U}(\omega, r, r'),
\label{cu}
\eeq
with
\bea
\overrightarrow{C}_{U}(\omega, r, r') &=& \sum_{l = 1}^{\infty} (2l+1)P_{l}(\hat{r} \cdot \hat{r}')\overrightarrow{R}^{(1)}_{\omega l}(r)\overrightarrow{R}^{(1 *)}_{\omega l}(r')\left(1 + \frac{1}{e^{2\pi\omega/\kappa}-1}\right)
\nn\\
\overleftarrow{C}_{U}(\omega, r, r') &=& \theta(\omega)\sum_{l = 1}^{\infty} (2l+1)P_{l}(\hat{r} \cdot \hat{r}')\overleftarrow{R}^{(1)}_{\omega l}(r)\overleftarrow{R}^{(1 *)}_{\omega l}(r').
\label{cu1}
\eea
The above integrals can be easily solved and we find that
\bea
&&\Biggl\langle \frac{d H_A}{dt} \Biggr\rangle_{VF} = -\frac{1}{16\pi^2}\sum_{\nu'}\sum_{k,j = 1}^{2}\,
{\cal U}^{rr}_{kj}(\nu ,\nu')\,\sqrt{g_{00}(r_k)g_{00}(r_j)}\,\exp\left[i\left(\widetilde{\Delta \nu}_k - \widetilde{\Delta \nu}_j\right)t\right]\,\Delta\nu
\nn\\
&\times&\int_{-\infty}^{\infty}\,d\omega\,\omega\,\Biggl\{C_{U}(\omega, r_k, r_j) \biggl[\frac{e^{i \Delta t/2 \left(\widetilde{\Delta \nu}_j -\omega\right)}\sin\left(\Delta t/2 \left(\widetilde{\Delta \nu}_j -\omega\right)\right)}{\widetilde{\Delta \nu}_j-\omega}
+\frac{e^{-i \Delta t/2 \left(\widetilde{\Delta \nu}_k -\omega\right)}\sin\left(\Delta t/2 \left(\widetilde{\Delta \nu}_k -\omega\right)\right)}{\widetilde{\Delta \nu}_k -\omega}\biggr] 
\nn\\
&+& C_{U}(\omega, r_j, r_k) \biggl[\frac{e^{i \Delta t/2 \left(\sqrt{g_{00}(r_j)}\,\Delta E +\omega\right)}\left(\sin\left(\Delta t/2 \left(\widetilde{\Delta \nu}_j +\omega\right)\right)\right)}{\widetilde{\Delta \nu}_j+\omega}
+ \frac{e^{-i \Delta t/2 \left(\widetilde{\Delta \nu}_k +\omega\right)}\sin\left(\Delta t/2 \left(\widetilde{\Delta \nu}_k +\omega\right)\right)}{\widetilde{\Delta \nu}_k+\omega}\biggr] \Biggr\}.
\nn\\
\label{vacuum-u}
\eea
As in the cases above studied, the atoms disentangle and can also entangle in a finite observation time due to vacuum fluctuations of the electromagnetic field. The creation of entanglement due to vacuum fluctuations also persists at late times.

Now let us present the radiation-reaction contributions. Performing similar calculations as above one gets:
\bea
&&\Biggl\langle \frac{d H_A}{dt} \Biggr\rangle_{RR} = -\frac{1}{16\pi^2}\sum_{\nu'}\sum_{k,j = 1}^{2}\,
{\cal U}^{rr}_{kj}(\nu ,\nu')\,\sqrt{g_{00}(r_k)g_{00}(r_j)}\,\exp\left[i\left(\widetilde{\Delta \nu}_k - \widetilde{\Delta \nu}_j\right)t\right]\,\Delta\nu
\nn\\
&\times&\int_{-\infty}^{\infty}\,d\omega\,\omega\,\Biggl\{C_{U}(\omega, r_k, r_j) \biggl[\frac{e^{i \Delta t/2 \left(\widetilde{\Delta \nu}_j -\omega\right)}\sin\left(\Delta t/2 \left(\widetilde{\Delta \nu}_j -\omega\right)\right)}{\widetilde{\Delta \nu}_j-\omega}
+\frac{e^{-i \Delta t/2 \left(\widetilde{\Delta \nu}_k -\omega\right)}\sin\left(\Delta t/2 \left(\widetilde{\Delta \nu}_k -\omega\right)\right)}{\widetilde{\Delta \nu}_k -\omega}\biggr] 
\nn\\
&-& C_{U}(\omega, r_j, r_k) \biggl[\frac{e^{i \Delta t/2 \left(\sqrt{g_{00}(r_j)}\,\Delta E +\omega\right)}\left(\sin\left(\Delta t/2 \left(\widetilde{\Delta \nu}_j +\omega\right)\right)\right)}{\widetilde{\Delta \nu}_j+\omega}
+ \frac{e^{-i \Delta t/2 \left(\widetilde{\Delta \nu}_k +\omega\right)}\sin\left(\Delta t/2 \left(\widetilde{\Delta \nu}_k +\omega\right)\right)}{\widetilde{\Delta \nu}_k+\omega}\biggr] \Biggr\}.
\nn\\
\label{radiation-u}
\eea
Observe that the effect of radiation reaction, with respect to absorption processes, do not contribute to the generation of entanglement between the atoms as in the cases investigated above; it always tend to disentangle an entangled state via spontaneous emission processes. As in the Hartle-Hawking case such a contribution is also altered by the appearance of a thermal contribution. 

For completeness, let us present the total rate of change of the atomic energy. This is obtained by adding the contributions of vacuum fluctuations and radiation reaction. One gets, for sufficiently large $\Delta t$:
\bea
\Biggl\langle \frac{d H_A}{dt} \Biggr\rangle_{tot} &=& -\frac{1}{8\pi}\sum_{\nu'}\sum_{k,j = 1}^{2}\,
{\cal U}^{rr}_{kj}(\nu ,\nu')\,\sqrt{g_{00}(r_k)g_{00}(r_j)}\,\exp\left[i\left(\widetilde{\Delta \nu}_k - \widetilde{\Delta \nu}_j\right)t\right]\,\Delta\nu
\nn\\
&\times&\, \biggl[\widetilde{\Delta \nu}_j\,C_{U}(\widetilde{\Delta \nu}_j, r_k, r_j)
+\widetilde{\Delta \nu}_k\,C_{U}(\widetilde{\Delta \nu}_k, r_k, r_j)\biggr].
\label{total-u}
\eea
\end{widetext}
As in the Hartle-Hawking case, the balance between vacuum fluctuations and radiation reaction which existed in the Boulware vacuum is disturbed and entanglement can be created via absorption process in the exterior region of the black hole even for large asymptotic $\Delta t$. The structure of the rate of variation of atomic energy implies the existence of thermal radiation from the black hole which is backscattered by space-time curvature. This thermal radiation is responsible for the possibility of creation of entanglement between atoms initially prepared in the ground state. The temperature of the thermal radiation as felt by each of the atoms is again given by equation~(\ref{t-haw}). Moreover, equation~(\ref{total-u}) shows us that the entangled state $|\Omega^{-}\rangle$ with the choice $c_1 = c_2 = 1/\sqrt{2}$ is stable with respect to radiative processes for atoms located at the same position.

Let us briefly digress on the rate of change of the atomic energy for the asymptotic regions of interest. Again we are assuming a large enough $\Delta t$ so that one could approximate $\sin x/x$ as delta functions. Consider the asymptotic region $r_1, r_2 \to \infty$. From the results derived in Appendix~\ref{B} one gets:
\bea
C_{U}(\omega, r_i, r_j) &\approx& H_{+}(\omega, r_i, r_j)\left(1 + \frac{1}{e^{2\pi\omega/\kappa}-1}\right) 
\nn\\
&+&\, \theta(\omega)\,F(\omega, {\bf x}_i,{\bf x}_j).
\eea
Note that thermal terms are multiplied by the gray-body factor $H_{\pm}$ which vanishes at spatial infinity. Hence as the atoms approach spatial infinity, the flux felt by them becomes more pale, which means that creation of entanglement by absorption processes becomes rarer.

The other important region is when $r_1, r_2 \to 2M$. One has that
\begin{widetext}
\bea
C_{U}(\omega, r_i, r_j) &\approx& \frac{16 a^2(r_i) \sinh(\pi \xi(\omega))}{g_{00}(r_i)\,\pi \xi(\omega)}
\left[A_{i\xi(\omega)}(\gamma, g_{00}(r_j), g_{00}(r_i)) + A_{-i\xi(\omega)}(\gamma, g_{00}(r_j), g_{00}(r_i))\right]\left(1 + \frac{1}{e^{2\pi\omega/\kappa}-1}\right)
\nn\\
&+&\,\theta(\omega)\,H_{-}(\omega, r_i, r_j).
\eea
Again note that $g_{00}(r)$ vanishes as the event horizon is approached, thence the rate of change of the atomic energy diverges. 

As a last analysis concerning the Unruh vacuum we take, say, $r_2 \to 2M$ whereas $r_1$ is kept arbitrary. One gets:
\bea
C_{U}(\omega, r_1, r_2) &\approx& \sum_{l=1}^{\infty}\frac{l(l+1) (2l + 1)\,P_{l}(\hat{r}_1 \cdot \hat{r}_2)}{(r_2)^2 (r_1)^2 \omega^2} 
\nn\\
&\times&\,\left[\theta(\omega)\overleftarrow{{\cal R}}^{(1)}_{\omega l}(r_1){\cal B}^{*}_{l}(\omega) e^{i\omega  r_{*2}} + \frac{2e^{-i\omega/2\kappa}}{\Gamma(i\omega/\kappa)} \overrightarrow{{\cal R}}^{(1)}_{\omega l}(r_1) e^{i\omega \ln l/\kappa} K_{i\omega/\kappa}\left(2l\sqrt{g_{00}(r_2)}\right)\left(1 + \frac{1}{e^{2\pi\omega/\kappa}-1}\right)\right].
\nn\\
\eea
\end{widetext}
From the results found in Appendix~\ref{B}, as $r_1 \to \infty$ the cross terms vanish and again we are left only with terms corresponding to isolated atoms. On the other hand, as $r_1$ approaches $2M$, $\langle d H_A/dt \rangle$ diverges. This last case confirms again the general result stated above: as the atoms approach the event horizon the rate of variation of atomic energy grows quickly. For large $\Delta t$ this implies a greatly enhanced generation of entanglement between the atoms initially prepared in the ground state.

\section{Conclusions and Perspectives}
\label{conclude}

Many works in the recent literature of quantum information theory have been devoted to investigations of entanglement in quantum field theory and quantum field theory in curved space-time. Throughout the text some of these were already cited. Among several investigations in the field, we would also like to mention the analysis regarding quantum teleportation between noninertial observers~\cite{als1,als2,shiw}, relativistic approaches to the Einstein-Podolsky-Rosen framework and also to Bell's inequality~\cite{cza,tera,kim,cab}. References~\cite{peres,ging,ging2,shi,cza2,jor,ade,ani} provide more intriguing discussions on the subject of relativistic quantum entanglement for the interested reader. The point is that most of these studies were implemented in a framework of open quantum systems. Employing the formalism developed by Dalibard, Dupont-Roc, and Cohen-Tannoudji, we have uncovered the distinct effects of vacuum fluctuations and radiation reaction on the quantum entanglement between two identical atoms in Schwarzschild space-time. Within such a formalism the interplay between vacuum fluctuations and radiation reaction can be considered to maintain the stability of the entangled state in some particular situations. We assume that both atoms are coupled to a quantum electromagnetic field. The overall picture is the following. The rate of change of the two-atom system energy is very small when the atoms are far away of the horizon. As they get closer, this rate increases in a oscillatory regime in such a way that, when the atoms approach the horizon, most contributions to the rate oscillates violently. This suggests that the generation of entanglement is highly magnified when the atoms are near the horizon and also largely suppressed when they get to spatial infinity. In turn, we have also obtained evidences that the degradation of entanglement follows the same response, i.e., it is highly enhanced when the atoms approach the horizon and also largely suppressed when they get to spatial infinity. The present analysis taken in connection with the results of references~\cite{china4,kempf} allow us to state the following assertion: Even though the thermal terms contribute decisively to the creation of entanglement between the atoms, the degree of entanglement thus generated is suppressed for atoms approaching the event horizon. In this way, we note that here the Hawking effect is a key ingredient in the discussion of creation of entanglement. We stress that one must not refrain from observing that the entanglement features of the system under consideration depend crucially on the distance of the atoms to the event horizon and also on the balance between vacuum fluctuations and radiation reaction. This is manifest in the framework studied here.

In this work since we are interested in mean lifes we choose an alternative perspective to understand quantum entanglement. We have carefully demonstrated that when considering the resonant interaction between two-level atoms, the machinery underlying entanglement can be understood as an interplay between classical concepts (represented by the radiation-reaction effect) and quantum-mechanical phenomena (vacuum fluctuations). In this scenario the usage of the DDC formalism has been proved to be essential in order to unfold this interpretation in a clear way. Nevertheless, we mention that the standard formalism for the evaluations of time evolution and correlation properties of entangled atomic systems is the traditional master equation approach. An important concept that is commonly addressed with the master equation approach is the entanglement swapping. Within this approach one derives an equation of motion for the reduced density operator of a certain subsystem A which interacts with another subsystem B. Commonly one describes the general solution of the master equation in terms of the so-called Kraus representation. In possession of a density matrix of the pair of atoms, it is possible to quantify the degree of entanglement by employing usual techniques, such as Wootters's concurrence or negativity. Concerning such entanglement measures, one could also consider the calculation of the entanglement entropy, which characterizes the correlations between sub-systems belonging to a quantum-mechanical system. A systematic study of entanglement entropy in quantum field theory and of black holes is given in may important works, see for instance Refs.~\cite{cardy,solodukhin} and references cited therein. We reserve future studies to all the important subjects raised above.

We believe that the results presented in this paper may have an impact in the studies of radiative process of atoms in the presence of an event horizon. A framework in which vacuum fluctuations and radiation-reaction effect have been clearly uncovered may contribute to a deeper understanding of such results. For instance, recently the method was employed to investigate the Casimir-Polder forces between two uniformly accelerated atoms~\cite{marino}. In such a work the authors exhibit a transition from the short distance thermal behavior dictated by the Unruh effect to a long distance nonthermal behavior. In addition, studies of quantum entanglement in Schwarzschild space-time are attracting much attention due to their obvious applications to the problem of black-hole information loss~\cite{hawking76,pres,gid,uald,hartle,nik}. One expects that the present investigation will impact the discussion on black-hole complementarity~\cite{suss} or even on the possible firewall scenarios~\cite{braun1,braun2,pol,verlinde}. Indeed, the relationship between particle detectors in different vacua in Rindler and Schwarzschild space-time was undertaken in recent studies~\cite{doug}. All these investigations suggest that the attempts to ascertain possible connections between the equivalence principle and quantum entanglement could unveil a different and important aspect on the black hole information paradox. Such subjects are under investigation and results will be reported elsewhere.

\section*{acknowlegements}

We thank N. F. Svaiter for useful discussions. Work supported by Conselho Nacional de Desenvolvimento Cientifico e Tecnol{\'o}gico do Brasil (CNPq).

\newpage

\appendix

\section{Correlation functions of the electromagnetic field in Schwarzschild space-time}
\label{A}

In this Appendix we present the correlation functions for each one of the vacuum states discussed above. A detailed analysis of the quantization of the electromagnetic field in Schwarzschild space-time can be found in reference~\cite{mat2}. For a different but related method see reference~\cite{cog}. The associated correlation functions are also evaluated in Ref.~\cite{china}. Fundamentally, the concept is to employ the modified Feynman gauge and then use the Gupta-Bleuler quantization in this gauge employing Schwarzschild coordinates. 

The action for the free electromagnetic fields in the modified Feynman gauge is given by
\beq
S_F = -\int d^4 x\,\sqrt{-g}\left[\frac{1}{4}\,F_{\alpha\beta}F^{\alpha\beta} + \frac{1}{2}\,G^2\right],
\label{ac-field}
\eeq
where $F_{\alpha\beta} = \nabla_{\alpha} A_{\beta} - \nabla_{\beta} A_{\alpha}$ and  $G = \nabla^{\mu}A_{\mu} + K^{\mu} A_{\mu}$, with $\nabla^{\mu}$ being the usual covariant derivative in curved space-time. Choosing:
$$
K^{\mu} = \left(0, \frac{2M}{r^2}, 0, 0\right),
$$
the equation for $A_0$ decouples from the other ones. A complete set of solutions of the field equations is denoted by $A^{(\lambda n; \omega l m)}_{\mu}$. The label $n$ distinguishes between modes incoming from the past null infinity ${\cal J}^{-}$ ($n = \leftarrow$) and those going out from the past horizon ${\cal H}^{-}$ ($n = \rightarrow$). There are four classes of modes which form this basis. The modes with $\lambda = 0$ do not obey the gauge condition $G = 0$ which is satisfied by all other modes with $\lambda \neq 0$. In turn, modes with $\lambda = 3$ are so-called pure gauge modes. Finally, the modes with $\lambda = 1, 2$ correspond to the physical modes. We choose them to have $A_0 = 0$. These are given by
\begin{widetext}
\beq
A^{(1 n; \omega l m)}_{\mu} = \left(0, R^{(1n)}_{\omega l}(r) Y_{lm}\,e^{-i\omega t}, \frac{(1 - 2M/r)}{l(l+1)}\frac{d}{dr}\left(r^2\,R^{(1n)}_{\omega l}(r)\right)\,\partial_{i}Y_{lm}\,e^{-i\omega t}\right)
\eeq
\end{widetext}
for $\lambda = 1$, with $i = \theta, \phi$ and $l \geq 1$. The functions $Y_{lm} = Y_{lm}(\theta, \phi)$ are the usual spherical harmonics and $l = 0, 1, 2, ...$, with $l \geq m$, $m$ being an integer number. As for $\lambda = 2$, they can be expressed as:
\beq
A^{(2 n; \omega l m)}_{\mu} = \left(0, 0, R^{(2n)}_{\omega l}(r) Y^{lm}_{i}\,e^{-i\omega t}\right),
\eeq
where the functions $Y^{lm}_{i} = Y^{lm}_{i}(\theta, \phi)$ are divergence-free vector spherical harmonics defined on the unit $2$-sphere with angular coordinates $(\theta, \phi)$. The associated normalization of the modes $A^{(\lambda n; \omega l m)}_{\mu}$ can be fixed from the usual inner product. Expanding the field operator in terms of the complete set of basic modes as
\bea
\hat{A}_{\mu}(x) &=& \sum_{\lambda n l m}\,\int_{0}^{\infty}\,\frac{d\omega}{\sqrt{4\pi\omega}}\biggl[A^{(\lambda n; \omega l m)}_{\mu}(x)\,\hat{a}^{(\lambda n)}_{\omega l m} 
\nn\\
&+&\, \left(A^{(\lambda n; \omega l m)}_{\mu}\right)^{*}(x)\,\hat{a}^{\dagger (\lambda n)}_{\omega l m}\biggr],
\label{boul}
\eea
The commutation relations between the annihilation and creation operators are given by
\bea
\left[\hat{a}^{(3 n)}_{\omega l m}, \hat{a}^{\dagger(3 n')}_{\omega' l' m'}\right] &=& -\left[\hat{a}^{(0 n)}_{\omega l m}, \hat{a}^{\dagger(3 n')}_{\omega' l' m'}\right] 
\nn\\
&=& \delta^{n n'}\delta_{ll'}\delta_{mm'}\delta(\omega - \omega')
\eea
and
\bea
\left[\hat{a}^{(1 n)}_{\omega l m}, \hat{a}^{\dagger(1 n')}_{\omega' l' m'}\right] &=& \left[\hat{a}^{(2 n)}_{\omega l m}, \hat{a}^{\dagger(2 n')}_{\omega' l' m'}\right] 
\nn\\
&=& \delta^{n n'}\delta_{ll'}\delta_{mm'}\delta(\omega - \omega').
\eea
All other commutators vanish. From the Gupta-Bleuler condition one gets $\hat{G}^{+} | \phi \rangle = 0$, for any physical state $| \phi \rangle$, where $\hat{G}^{+}$ is the positive-frequency part of the operator $\hat{G} = \nabla^{\mu}\hat{A}_{\mu} + K^{\mu} \hat{A}_{\mu}$. This condition means that any state with $\lambda = 3$ is unphysical and the states with $\lambda = 0$ have zero norm and are orthogonal to any physical states. The Boulware vacuum $|0_B\rangle$~\cite{boul} is defined by requiring that it be annihilated by all annihilation operators $\hat{a}^{\dagger(\lambda n)}_{\omega l m}$ . One can take as the representative elements the states obtained by applying the creation operators $\hat{a}^{\dagger(3 n)}_{\omega l m}$, $\lambda = 1, 2$ on the vacuum $|0_B\rangle$. As discussed in reference~\cite{mat2}, unphysical particles created by $\hat{a}^{\dagger(3 n)}_{\omega l m}$ will be in thermal equilibrium in the so-called Hartle-Hawking vacuum~\cite{haw} for a static black hole if one demands that the gauge-fixed two-point function to be non-singular on the horizons similar to the procedure taken in the scalar-field case~\cite{wald}. There will also be a flux of unphysical particles in the so-called Unruh vacuum~\cite{unruh}. 

Observe that the positive-frequency states defined as above are related to the timelike Killing vector field $\partial / \partial t$ with respect to which the exterior region of the black hole is static~\cite{sciama}. However, as argued in references~\cite{unruh,sciama}, there are other possible prescriptions when considering the metric in Kruskal-Szekeres coordinates instead of the usual Schwarzschild coordinates. In this respect, one easily notes that the null coordinates on the past horizon  ${\cal H}^{-}$ ($U = v - u$) and the null coordinate on the future horizon ${\cal H}^{+}$ ($V = v + u$) also act as Killing vector fields $\partial / \partial U$ and $\partial / \partial V$ on the respective horizons. Therefore, one can define basis modes in terms of such Kruskal null coordinates $U, V$. This set is regular on the entire manifold. The associated vacuum $|0_H\rangle$ is known as the Hartle-Hawking vacuum. On the other hand, it is also known that one may take the incoming modes to be positive frequency with respect to $\partial / \partial t$ and the outgoing modes to be positive frequency with respect to $\partial / \partial U$ -- one can show that such a prescription leads to a definition of a set of modes which oscillate infinitely rapidly on the past event horizon~\cite{birrel}. The associated vacuum $|0_U\rangle$ is known as the Unruh vacuum~\cite{unruh}. This last prescription is the one required in order to mock up the geometrical effects associated with the gravitational collapse of a spherically symmetric electrically neutral star.

Let us present the expansion of the field operator in terms of the complete set of modes associated with the Hartle-Hawking vacuum and the Unruh vacuum. Our discussion has grounds on the reference~\cite{unruh}. As discussed in such a reference, the field operators can also be expanded as
\begin{widetext}
\bea
\hat{A}_{\mu}(x) &=& \sum_{\lambda l m}\,\Biggl\{\int_{-\infty}^{\infty}\,d\omega\,\frac{1}{\sqrt{8\pi\omega\sinh(4\pi M \omega)}}\left[\bar{A}^{(\lambda \leftarrow; \omega l m)}_{\mu}(x)\,\hat{h}^{(\lambda \leftarrow)}_{\omega l m} + \left(\bar{A}^{(\lambda \leftarrow; \omega l m)}_{\mu}\right)^{*}(x)\,\hat{h}^{\dagger (\lambda \leftarrow)}_{\omega l m}\right]
\nn\\
&+& \int_{-\infty}^{\infty}d\omega\,\frac{1}{\sqrt{8\pi\omega\sinh(4\pi M \omega)}}\left[\bar{A}^{(\lambda \rightarrow; \omega l m)}_{\mu}(x)\,\hat{h}^{(\lambda \rightarrow)}_{\omega l m} + \left(\bar{A}^{(\lambda \rightarrow; \omega l m)}_{\mu}\right)^{*}(x)\,\hat{h}^{\dagger (\lambda \rightarrow)}_{\omega l m}\right]\Biggr\},
\label{haw}
\eea
where
\bea
 \bar{A}^{(\lambda \rightarrow; \omega l m)}_{\mu}(x) &=& e^{2\pi M \omega}A^{(\lambda \rightarrow; \omega l m)}_{\mu \textrm{I}}(x) + e^{-2\pi M \omega}\left(A^{(\lambda \rightarrow ; \omega l m)}_{\mu \textrm{III}}\right)^{*}(x)
\nn\\
 \bar{A}^{(\lambda \leftarrow; \omega l m)}_{\mu}(x) &=& e^{-2\pi M \omega}\left(A^{(\lambda \leftarrow ; \omega l m)}_{\mu \textrm{I}}\right)^{*}(x) + e^{2\pi M \omega}A^{(\lambda \leftarrow ; \omega l m)}_{\mu \textrm{III}}(x)
\eea
with $\bar{A}^{(\lambda n ; \omega l m)}_{\mu \textrm{I}}(x) = A^{(\lambda n; \omega l m)}_{\mu}(x)$ for $x \in \textrm{I}$ and zero for $x \in \textrm{III}$, I and III different regions of the Kruskal-Szekeres diagram as commented above [see figure ($31.3$) of Ref.~\cite{wheeler}], and similarly for $\bar{A}^{(\lambda n ; \omega l m)}_{\mu \textrm{III}}(x)$ which is zero for $x \in \textrm{I}$. One has
\bea
\left[\hat{h}^{(3 n)}_{\omega l m}, \hat{h}^{\dagger(3 n')}_{\omega' l' m'}\right] &=& -\left[\hat{h}^{(0 n)}_{\omega l m}, \hat{h}^{\dagger(3 n')}_{\omega' l' m'}\right] = \delta^{n n'}\delta_{ll'}\delta_{mm'}\delta(\omega - \omega')
\nn\\
\left[\hat{h}^{(1 n)}_{\omega l m}, \hat{h}^{\dagger(1 n')}_{\omega' l' m'}\right] &=& \left[\hat{h}^{(2 n)}_{\omega l m}, \hat{h}^{\dagger(2 n')}_{\omega' l' m'}\right] = \delta^{n n'}\delta_{ll'}\delta_{mm'}\delta(\omega - \omega').
\eea
with all other commutators vanishing, and also $\hat{h}^{(\lambda n)}_{\omega l m} |0_H\rangle = 0$. In turn, one may also expand the field operators as
\bea
\hat{A}_{\mu}(x) &=& \sum_{\lambda l m}\,\Biggl\{\int_{0}^{\infty}\,\frac{d\omega}{\sqrt{4\pi\omega}}\left[A^{(\lambda \leftarrow; \omega l m)}_{\mu}(x)\,\hat{u}^{(\lambda \leftarrow)}_{\omega l m} + \left(A^{(\lambda \leftarrow; \omega l m)}_{\mu}\right)^{*}(x)\,\hat{u}^{\dagger (\lambda \leftarrow)}_{\omega l m}\right]
\nn\\
&+& \int_{-\infty}^{\infty}d\omega\,\frac{1}{\sqrt{8\pi\omega\sinh(4\pi M \omega)}}\left[\bar{A}^{(\lambda \rightarrow; \omega l m)}_{\mu}(x)\,\hat{u}^{(\lambda \rightarrow)}_{\omega l m} + \left(\bar{A}^{(\lambda \rightarrow; \omega l m)}_{\mu}\right)^{*}(x)\,\hat{u}^{\dagger (\lambda \rightarrow)}_{\omega l m}\right]\Biggr\},
\label{un}
\eea
with
\bea
\left[\hat{u}^{(3 n)}_{\omega l m}, \hat{u}^{\dagger(3 n')}_{\omega' l' m'}\right] &=& -\left[\hat{u}^{(0 n)}_{\omega l m}, \hat{u}^{\dagger(3 n')}_{\omega' l' m'}\right] = \delta^{n n'}\delta_{ll'}\delta_{mm'}\delta(\omega - \omega')
\nn\\
\left[\hat{u}^{(1 n)}_{\omega l m}, \hat{u}^{\dagger(1 n')}_{\omega' l' m'}\right] &=& \left[\hat{u}^{(2 n)}_{\omega l m}, \hat{u}^{\dagger(2 n')}_{\omega' l' m'}\right] = \delta^{n n'}\delta_{ll'}\delta_{mm'}\delta(\omega - \omega').
\eea
\end{widetext}
with all other commutators vanishing, and $\hat{u}^{(\lambda n)}_{\omega l m} |0_U\rangle = 0$. For simplicity, we assume that the atoms are polarized along the radial direction defined by their positions relative to the black-hole space-time rotational Killing vector fields. This assumption significantly simplifies the calculations in that the contributions associated with the polarizations in angular directions does not need to be considered. Therefore, with the usual relationships $E_{i} = F_{0i}$, one can calculate the various correlation functions which will be important in our calculations. The important object to be considered is
$$
E_{r} = F_{0r} = \nabla_{0} A_{r} - \nabla_{r} A_{0} = \partial_{0} A_{r} - \partial_{r} A_{0}.
$$
(the connection terms cancel) Hence
\beq
\langle 0|\hat{E}_{r}(x)\hat{E}_{r}(x')|0\rangle = \langle 0|(\partial_{0} \hat{A}_{r} - \partial_{r} \hat{A}_{0})(\partial_{0}' \hat{A}_{r} - \partial_{r}' \hat{A}_{0})|0\rangle.
\eeq
Let us present the correlation functions for each one of the vacuum states for $x, x' \in \textrm{I}$.

\begin{widetext}
\begin{enumerate}

\item The Boulware vacuum 

One has:
\bea
\langle 0_B|\hat{E}_{r}(x)\hat{E}_{r}(x')|0_B\rangle &=& \frac{1}{4\pi}\sum_{l m}\,\int_{0}^{\infty}\,d\omega\,\omega\,e^{-i\omega (t-t')} Y_{lm}(\theta,\phi)Y^{*}_{lm}(\theta',\phi')
\nn\\
&\times&\,\left[\overrightarrow{R}^{(1)}_{\omega l}(r)\overrightarrow{R}^{(1 *)}_{\omega l}(r') + \overleftarrow{R}^{(1)}_{\omega l}(r)\overleftarrow{R}^{(1 *)}_{\omega l}(r')\right].
\eea
where we have used that
\beq
\langle 0_B|\hat{a}^{(1 n)}_{\omega l m}\hat{a}^{\dagger(1 n')}_{\omega' l' m'}|0_B\rangle = \delta^{n n'}\delta_{ll'}\delta_{mm'}\delta(\omega - \omega'),
\label{comb}
\eeq
and all other possible combinations coming from the product $\hat{E}_{r}(x)\hat{E}_{r}(x')$ vanish.

\item The Hartle-Hawking vacuum

One has
\bea
\langle 0_H|\hat{E}_{r}(x)\hat{E}_{r}(x')|0_H\rangle &=& \frac{1}{4\pi}\sum_{l m}\,\int_{-\infty}^{\infty}\,d\omega\,\omega\,\Biggl[e^{-i\omega (t-t')} Y_{lm}(\theta,\phi)Y^{*}_{lm}(\theta',\phi')\frac{\overrightarrow{R}^{(1)}_{\omega l}(r)\overrightarrow{R}^{(1 *)}_{\omega l}(r')}{1 - e^{-2\pi\omega/\kappa}}
\nn\\
&+& e^{i\omega (t-t')} Y^{*}_{lm}(\theta,\phi)Y_{lm}(\theta',\phi')\frac{\overleftarrow{R}^{(1*)}_{\omega l}(r)\overleftarrow{R}^{(1)}_{\omega l}(r')}{e^{2\pi\omega/\kappa} - 1}\Biggr],
\eea
where $\kappa = 1/4M$ is the surface gravity of the black hole. A relation similar to~(\ref{comb}) holds for the operators $\hat{h}^{(1 n)}_{\omega l m}$, $\hat{h}^{\dagger(1 n)}_{\omega l m}$.\\

\item The Unruh vacuum

One has:
\bea
\langle 0_U|\hat{E}_{r}(x)\hat{E}_{r}(x')|0_U\rangle &=& \frac{1}{4\pi}\sum_{l m}\,\int_{-\infty}^{\infty}\,d\omega\,\omega\,e^{-i\omega (t-t')} Y_{lm}(\theta,\phi)Y^{*}_{lm}(\theta',\phi')
\nn\\
&\times&\,\left[\frac{\overrightarrow{R}^{(1)}_{\omega l}(r)\overrightarrow{R}^{(1 *)}_{\omega l}(r')}{1 - e^{-2\pi\omega/\kappa}} + \theta(\omega)\overleftarrow{R}^{(1)}_{\omega l}(r)\overleftarrow{R}^{(1 *)}_{\omega l}(r')\right],
\eea
where $\theta(z)$ is the usual Heaviside theta function. A relation similar to~(\ref{comb}) also holds for the operators $\hat{u}^{(1 n)}_{\omega l m}$, $\hat{u}^{\dagger(1 n)}_{\omega l m}$.

\end{enumerate}

\end{widetext}

We stress that each of the vacua discussed above represent different physical scenarios. The  Boulware state, empty at infinity,
is the ground state of quantum fields around a cold neutron star. The Unruh state, empty at past null infinity and regular on the future horizon, is customarily taken as ground state for an evaporating black hole, i.e., such a vacuum state reproduces the effects of a gravitational collapsing body. In turn, the Hartle-Hawking vacuum corresponds to a black hole in equilibrium with an infinite sea of black-body radiation.

\section{Evaluation of mode sums}
\label{B}

In order to evaluate the correlation functions one needs to present explicit expressions for the radial functions. Even though it is a remarkable task, fortunately one is usually interested in two asymptotic regions, namely $r \to 2M$ (near the event horizon) and $r \to \infty$ (away from the event horizon). In this case, the behavior of the radial functions is well known. We shall briefly discuss such limits in the present Appendix. We extend the results of Refs.~\cite{candelas,china} for correlation functions calculated at different points of the space-time. From standard considerations, one has that
\bea
\overrightarrow{{\cal R}}^{(1)}_{\omega l}(r) &\sim& e^{i\omega r_{*}} + \overrightarrow{{\cal A}}_{l}(\omega)e^{-i\omega r_{*}},\,\,\,r \to 2M
\nn\\
\overrightarrow{{\cal R}}^{(1)}_{\omega l}(r) &\sim& \overrightarrow{{\cal B}}_{l}(\omega)e^{i\omega r_{*}},\,\,\,r \to \infty
\nn\\
\overleftarrow{{\cal R}}^{(1)}_{\omega l}(r) &\sim& \overleftarrow{{\cal B}}_{l}(\omega)e^{-i\omega r_{*}},\,\,\,r \to 2M
\nn\\
\overleftarrow{{\cal R}}^{(1)}_{\omega l}(r) &\sim& e^{-i\omega r_{*}} + \overleftarrow{{\cal A}}_{l}(\omega)e^{i\omega r_{*}},\,\,\,r \to \infty
\label{asymp}
\eea
where $r_{*} = r + 2M\ln(r/2M - 1)$ is the Regge-Wheeler tortoise coordinate and ${\cal R}^{(1n)}_{\omega l}(r)$ is defined through the equation:
$$
R^{(1n)}_{\omega l}(r) = \frac{\sqrt{l(l+1)}}{\omega}\,\frac{{\cal R}^{(1n)}_{\omega l}(r)}{r^2}
$$
In the above ${\cal A}$ and ${\cal B}$ are the usual reflection and transmission coefficients, respectively, with the following properties
\bea
\overrightarrow{{\cal B}}_{l}(\omega) &=& \overleftarrow{{\cal B}}_{l}(\omega) = {\cal B}_{l}(\omega)
\nn\\
|\overrightarrow{{\cal A}}_{l}(\omega)| &=& |\overleftarrow{{\cal A}}_{l}(\omega)|
\nn\\
1 - |\overrightarrow{{\cal A}}_{l}(\omega)|^2 &=& 1 - |\overleftarrow{{\cal A}}_{l}(\omega)|^2 = |{\cal B}_{l}(\omega)|^2
\nn\\
\overrightarrow{{\cal A}}_{l}^{*}(\omega){\cal B}_{l}(\omega) &=& - {\cal B}^{*}_{l}(\omega)\overleftarrow{{\cal A}}_{l}(\omega)
\eea
Key results involving the mode summations in the asymptotic regions $r \to 2M$ and $r \to \infty$ will be now considered. At fixed radial distances $r$ and $r'$ the radial correlation function of the field in the Boulware vacuum is given by
\begin{widetext}
\beq
\langle 0_B|\hat{E}_{r}(x)\hat{E}_{r}(x')|0_B\rangle = \frac{1}{16\pi^2}\sum_{l=1}^{\infty}\,\int_{0}^{\infty}\,d\omega\,\omega\,e^{-i\omega (t-t')} (2l + 1)\,P_{l}(\hat{r} \cdot \hat{r}')
\left[\overrightarrow{R}^{(1)}_{\omega l}(r)\overrightarrow{R}^{(1 *)}_{\omega l}(r') + \overleftarrow{R}^{(1)}_{\omega l}(r)\overleftarrow{R}^{(1 *)}_{\omega l}(r')\right]
\label{b-sph}
\eeq
where we have used the addition theorem for the spherical harmonics~\cite{hilbert}
$$
\frac{4\pi}{2l + 1}\sum_{m = - l}^{l}Y_{lm}(\theta,\phi)Y^{*}_{lm}(\theta',\phi') = P_{l}(\hat{r} \cdot \hat{r}')
$$
where $\hat{r}$ and $\hat{r}'$ are two unit vectors with spherical coordinates $(\theta, \phi)$ and $(\theta', \phi')$, respectively, and $P_{l}$ is the Legendre polynomial of degree $l$~\cite{abram}. From equation~(\ref{asymp}) one has that
\beq
\sum_{l=1}^{\infty}(2l + 1)\,P_{l}(\hat{r} \cdot \hat{r}')\overrightarrow{R}^{(1)}_{\omega l}(r)\overrightarrow{R}^{(1 *)}_{\omega l}(r') \approx \sum_{l=1}^{\infty}\frac{l(l+1)(2l + 1)\,P_{l}(\hat{r} \cdot \hat{r}')|{\cal B}_{l}(\omega)|^2 e^{i\omega (r_{*} - r'_{*})}}{\omega^2 r^2 r'^2},\,\,\, r,r' \to \infty.
\label{asymp-inf}
\eeq
For ${\bf x} = {\bf x}'$ one gets ($P_{l}(1) = 1$):
\beq
\sum_{l=1}^{\infty} (2l + 1)|\overrightarrow{R}_{\omega l}|^2 \approx \sum_{l=1}^{\infty}\frac{l(l+1)(2l+1)|{\cal B}_{l}(\omega)|^2}{r^4\omega^2},\,\,\, r \to \infty.
\label{asymp-inf2}
\eeq
In turn, in order to estimate the remaining sum, one should note that the above correlation function at large radii should agree with the correlation function of the electric field in the Minkowski vacuum [a similar consideration was undertaken in reference~\cite{china}]. The latter is given by~\cite{cohen}
\bea
\langle 0 |\hat{E}^{i}(x)\hat{E}^{j}(x')| 0\rangle &=& \left(\frac{\partial}{\partial t}\frac{\partial}{\partial t'}\delta^{ij} - \frac{\partial}{\partial x_i}\frac{\partial}{\partial x_j'}\right)\,D(t-t',{\bf x} - {\bf x}'),
\nn\\
&=& -\left(\frac{\partial^2}{\partial \eta^2}\delta^{ij} - \frac{\partial}{\partial \rho_i}\frac{\partial}{\partial \rho_j}\right)\,D(\eta,\mbox{\small\mathversion{bold}${\rho}$}),
\eea
where $\eta = t - t'$, $\rho_i = ({\bf x} - {\bf x}')_{i}$ and
\bea
D(t - t',{\bf x} - {\bf x}') &=& \frac{1}{(2\pi)^3}\int \frac{d^3k}{2\omega_{{\bf k}}}\,e^{i [{\bf k}\cdot ({\bf x} - {\bf x}')-\omega_{{\bf k}}(t-t')]}  
\nn\\
&=& \frac{1}{(2\pi)^2}\int_{0}^{\infty} d\omega\,e^{-i\omega(t-t')}\,\frac{\sin\left(\omega|{\bf x} - {\bf x}'|\right)}{|{\bf x} - {\bf x}'|},
\eea
with $\omega_{{\bf k}} = \omega = |{\bf k}|$. Performing the derivatives, one gets, with $\Delta t = t - t'$ and $\Delta{\bf x} = {\bf x} - {\bf x}'$:
\beq
\langle 0 |\hat{E}^{i}(x)\hat{E}^{j}(x')| 0\rangle = \frac{1}{16\pi^2}\int_{0}^{\infty}\,d\omega\,\omega\,e^{-i\omega \Delta t}\,D^{ij}(\omega, {\bf x},{\bf x}'),
\label{m-car}
\eeq
where
\beq
D_{ij}(\omega, {\bf x},{\bf x}') = - \frac{4}{\omega|\Delta{\bf x}|^3}\left[\delta_{ij}{\cal S}_{1}(\omega, |\Delta{\bf x}|) - \frac{(\Delta{\bf x})_{i}(\Delta{\bf x})_{j}}{|\Delta{\bf x}|^2}\,{\cal S}_{3}(\omega, |\Delta{\bf x}|)\right],
\eeq
with
$$
{\cal S}_{n}(\omega, |\Delta{\bf x}|) = \left(n - \omega^2|\Delta{\bf x}|^2\right)\sin\left(\omega|\Delta{\bf x}|\right) - n\omega|\Delta{\bf x}|\cos\left(\omega|\Delta{\bf x}|\right),
$$
and we have used that
$$
\frac{\partial}{\partial \rho_i}f(\rho) = \frac{\rho^i}{\rho}\frac{d}{d\rho}f(\rho),
$$
$$
\frac{\partial}{\partial \rho_i}\frac{\partial}{\partial \rho_j}f(\rho) = \frac{\delta^{ij}}{\rho}\frac{d}{d\rho}f(\rho) + \frac{\rho^i \rho^{j}}{\rho}\frac{d}{d\rho}\left[\frac{1}{\rho}\frac{d}{d\rho}f(\rho)\right].
$$
In order to compare equations~(\ref{b-sph}) and~(\ref{m-car}), the latter should be expressed in spherical coordinates. This can be achieved using the usual transformation formula between the Cartesian unit vectors and the Spherical unit vectors, which leads us to:
$$
\langle 0 |\hat{E}_{i}(y)\hat{E}_{j}(y')| 0\rangle = \frac{\partial x^{a}}{\partial y^{i}}\frac{\partial x^{'b}}{\partial y^{'j}}\langle 0 |\hat{E}_{a}(x)\hat{E}_{b}(x')| 0\rangle,
$$
where the $y_j$ are the usual spherical coordinates $r, \theta, \phi$. Hence
\beq
\langle 0 |\hat{E}_{r}(y)\hat{E}_{r}(y')| 0\rangle = \frac{1}{16\pi^2}\int_{0}^{\infty}\,d\omega\,\omega\,e^{-i\omega \Delta t}
F(\omega, {\bf x},{\bf x}'),
\label{m-sph}
\eeq
where
\bea
F(\omega, {\bf x},{\bf x}') &=& \sin\theta\sin\theta'\left[\cos\phi\cos\phi' D_{11}(\omega, {\bf x},{\bf x}') + \sin\phi\sin\phi' D_{22}(\omega, {\bf x},{\bf x}') \right] + \cos\theta\cos\theta' D_{33}(\omega, {\bf x},{\bf x}') 
\nn\\
&+&\, \left(\cos\theta\sin\theta'\cos\phi' + \cos\theta'\sin\theta\cos\phi\right)D_{13}(\omega, {\bf x},{\bf x}') 
\nn\\
&+&\, \left(\cos\theta\sin\theta'\sin\phi'  + \cos\theta'\sin\theta\sin\phi\right)D_{23}(\omega, {\bf x},{\bf x}')
\nn\\
&+&\, \sin\theta\sin\theta'\sin(\phi + \phi')D_{12}(\omega, {\bf x},{\bf x}'),
\label{f}
\eea
with ${\bf x},{\bf x}'$ expressed in spherical coordinates. Therefore, comparing equations~(\ref{b-sph}) and~(\ref{m-sph}) one gets, for $r, r' \to \infty$:
\beq
\sum_{l=1}^{\infty}(2l + 1)\,P_{l}(\hat{r} \cdot \hat{r}')\overleftarrow{R}^{(1)}_{\omega l}(r)\overleftarrow{R}^{(1 *)}_{\omega l}(r') \approx F(\omega, {\bf x},{\bf x}'),\,\,\, r, r' \to \infty.
\label{asymp-inf3}
\eeq
For ${\bf x} = {\bf x}'$:
\beq
\sum_{l=1}^{\infty} (2l + 1)|\overleftarrow{R}_{\omega l}|^2 \approx \frac{8\omega^2}{3},\,\,\, r \to \infty.
\label{asymp-inf4}
\eeq
In order to evaluate the mode sums in the region $r \sim 2M$ a certain amount of caution is mandatory. We begin by defining
$$
\zeta^2 = \frac{r}{2M} - 1,
$$
and
$$
\xi = 4 M \omega.
$$
With these definitions, and using that $l(l+1)\zeta^2 \approx (l\zeta)^2$ (since $\zeta \sim 0$), one can easily prove that $\overrightarrow{{\cal R}}^{(1)}_{\omega l}$, taken as a function of $\zeta$, obeys the following differential equation 
\beq
\left[\zeta^2\frac{d^2}{d\zeta^2} + \zeta\frac{d}{d\zeta} + \left(\xi^2 - (2l\zeta)^2\right)\right]\overrightarrow{{\cal R}}^{(1)}_{\omega l}(\zeta) = 0,
\eeq
whose solutions are the modified Bessel functions $K_{i\xi}(2l\zeta)$ and $I_{i\xi}(2l\zeta)$. Whence the general solution can be conveniently expressed as:
\beq
\overrightarrow{{\cal R}}^{(1)}_{\omega l}\big|_{r \to 2M} \approx c_{l} K_{i\xi}(2l\zeta) + d_{l} I_{-i\xi}(2l\zeta) . 
\label{sol-varphi}
\eeq
As $l \to \infty$ for fixed $\zeta$, the function $\overrightarrow{{\cal R}}^{(1)}_{\omega l} \to 0$, since $r$ lies then in the region for which the effective potential for the radial function is large. One deduces from this that $d_{l}$ is an exponentially small function of $l$ for large $l$ since~\cite{abram}
$$
I_{\nu}(z) \sim \frac{e^{z}}{\sqrt{2\pi z}},
$$
which is valid for large $z$ and fixed $\nu$. The second term in equation~(\ref{sol-varphi}) will therefore make a contribution to the sum in equation~(\ref{b-sph}) which remains bounded as $\zeta \to 0$ and which may be neglected in comparison with that of the first term in~(\ref{sol-varphi}) which will be of order $(\zeta\zeta')^{-2}$. The coefficient $c_{l}$ may be determined by comparing the result~\cite{abram}
$$
K_{\nu}(z) \sim \frac{1}{2}\,\Gamma(\nu) \left(\frac{z}{2}\right)^{-\nu}
$$
($\Gamma(z)$ is the usual gamma function) which is valid for $\Re[\nu] > 0$ fixed and $z \to 0$ with the asymptotic solution
$$
\overrightarrow{{\cal R}}^{(1)}_{\omega l}(r) \sim e^{i\omega r_{*}} + \overrightarrow{{\cal A}}_{l}(\omega)e^{-i\omega r_{*}},\,\,\,r \to 2M.
$$
One finds that
\beq
c_{l} \sim \frac{2e^{i\xi/2}\,l^{-i\xi}}{\Gamma(-i\xi)}.
\eeq
Hence, to leading order
\bea
&&\sum_{l=1}^{\infty}(2l + 1)\,P_{l}(\hat{r} \cdot \hat{r}')\overrightarrow{R}^{(1)}_{\omega l}(r)\overrightarrow{R}^{(1 *)}_{\omega l}(r') \approx \frac{4}{\Gamma(i\xi)\Gamma(-i\xi)}\sum_{l=1}^{\infty}\frac{l(l+1)(2l + 1)\,P_{l}(\cos\gamma)}{\omega^2 r^2 r'^2}\,K_{i\xi}(2l\zeta)K_{i\xi}(2l\zeta')
\nn\\
&&\approx \frac{2\sinh(4\pi M \omega)}{\pi M^3 \omega}\int_{0}^{\infty}dl\,l^3\,J_{0}(l\gamma)\,K_{i\xi}\left(2l\sqrt{g_{00}(r)}\right)K_{i\xi}\left(2l\sqrt{g_{00}(r')}\right),\,\,\, r, r' \to 2M
\eea
where $g_{00} = (1 - 2M/r)$, $\cos\gamma = \hat{r} \cdot \hat{r}' = \cos\theta\cos\theta' + \sin\theta\sin\theta'\cos(\phi - \phi')$ and we have used that~\cite{abram}
$$
\Gamma(i\xi)\Gamma(-i\xi) = \frac{\pi}{\xi\sinh(\pi \xi)},
$$
together with the asymptotic result:
$$
P_{\nu}\left(\cos\frac{x}{\nu}\right) \approx J_{0}(x) + {\cal O}(\nu^{-1}),
$$
in which $J_{\mu}(x)$ is a Bessel function of the first kind. Employing the result~\cite{prud}
\beq
\int_{0}^{\infty}dx\, x^{\alpha - 1}\,J_{\lambda}(ax) K_{\mu}(2\sqrt{b}x) K_{\nu}(2\sqrt{c}x) = \frac{2^{\alpha - 3}a^{\lambda}}{(2\sqrt{c})^{\alpha + \lambda}\Gamma(\lambda + 1)}\left[A^{\nu}_{\mu}(a, b, c) + A^{\nu}_{-\mu}(a, b, c)\right]
\eeq
where ($A^{\mu}_{\mu} = A_{\mu}$)
\bea
A^{\nu}_{\mu}(a, b, c) &=& \left(\frac{b}{c}\right)^{\mu/2}\Gamma\left[-\mu, \frac{\alpha + \lambda + \mu - \nu}{2}, \frac{\alpha + \lambda + \mu + \nu}{2}\right]
\nn\\
&\times&\,F_{4}\left(\frac{\alpha + \lambda + \mu - \nu}{2}, \frac{\alpha + \lambda + \mu + \nu}{2}; \lambda + 1, \mu + 1; -\frac{a^2}{4c}, \frac{b}{c}\right),
\eea
$F_4(a, b; c, c'; x, y)$ being the Appell Hypergeometric Function $F_4$
$$
F_4(a, b; c, c'; x, y) = \sum_{m, n = 0}^{\infty}\,\frac{(a)_{m+n}(b)_{m+n}}{(c)_{m}(c')_{n} m! n!}\,x^{m} y^{n}
$$
($(q)_{m}$ is the Pochhammer symbol representing the rising factorial) and also 
$$\Gamma[a_1, \cdots, a_m] = \prod_{k=1}^{m}\Gamma(a_k),$$
one gets:
\bea
&&\sum_{l=1}^{\infty}(2l + 1)\,P_{l}(\hat{r} \cdot \hat{r}')\overrightarrow{R}^{(1)}_{\omega l}(r)\overrightarrow{R}^{(1 *)}_{\omega l}(r') \approx \frac{\sinh(4\pi M \omega)}{4\pi g^{2}_{00}(r) M^3 \omega}
\nn\\
&&\times\,\left[A_{4M\omega i}(\gamma, g_{00}(r'), g_{00}(r)) + A_{-4M\omega i}(\gamma, g_{00}(r'), g_{00}(r))\right],\,\,\, r, r' \to 2M.
\label{asymp-2m}
\eea
One may derive a much simpler result by considering that $r \approx r'$ but ${\bf \hat{r}} \neq {\bf \hat{r}}'$. With~\cite{prud}:
\bea
\int_{0}^{\infty}dx\,x^3 J_0(a x) [K_{i q}(2 b x)]^2 &=& \frac{4 \pi  \text{csch}(\pi  q)}{a^4 \left(a^2+16 b^2\right)^2 \sqrt{\frac{16 b^2}{a^2}+1}} 
\nn\\
&\times&\,\Biggl\{2 a^2 q \sqrt{\frac{16 b^2}{a^2}+1} \left(a^2+4 b^2\right) \cos \left[2 q \text{csch}^{-1}\left(\frac{4 b}{a}\right)\right]
\nn\\
&+&\, \left(a^4 \left(q^2-1\right)+8 a^2 b^2 \left(2 q^2-1\right)-64 b^4\right) \sin \left[2 q \text{csch}^{-1}\left(\frac{4 b}{a}\right)\right]\Biggr\},
\eea
where we assume a small positive imaginary part for $a$ so that the integral converges. One gets
\bea
&&\sum_{l=1}^{\infty}(2l + 1)\,P_{l}(\hat{r} \cdot \hat{r}')\overrightarrow{R}^{(1)}_{\omega l}(r)\overrightarrow{R}^{(1 *)}_{\omega l}(r') \approx 
\frac{4 \pi  \text{csch}(\pi\xi)}{\gamma^3 \left(16 g_{00}(r)+\gamma^2\right)^{5/2}} 
\nn\\
&\times&\,\Biggl\{2\gamma \xi\left[16 g_{00}(r)+\gamma^2\right]^{1/2} \left[4 g_{00}(r) + \gamma^2\right] 
\cos\left[2\xi \text{csch}^{-1}\left(\frac{4\sqrt{g_{00}(r)}}{\gamma}\right)\right]
\nn\\
&+&\, \left[\gamma^4 \left(\xi^2-1\right) + 8\gamma^2 g_{00}(r) \left(2 \xi^2-1\right) - 64 (g_{00}(r))^2\right] 
\sin \left[2\xi \text{csch}^{-1}\left(\frac{4\sqrt{g_{00}(r)}}{\gamma}\right)\right]\Biggr\},\,\,\, r, r' \to 2M.
\label{asymp-2m-2}
\eea
For ${\bf x} = {\bf x}'$:
\beq
\sum_{l=1}^{\infty}(2l + 1)\,|\overrightarrow{R}^{(1)}_{\omega l}|^2 \approx \frac{8\omega^2}{3g_{00}^2} + \frac{1}{6M^2 g_{00}^2},\,\,\, r \to 2M,
\label{asymp-2m2}
\eeq
where we have used that~\cite{prud}
$$
\frac{8}{\Gamma(i\xi)\Gamma(-i\xi)}\int_{0}^{\infty}dt\,t^3\,[K_{i\xi}\left(2tx\right)]^2 = \frac{\xi^2 \left(\xi^2+1\right)}{6 x^4}.
$$
The other mode sum in the region $r \sim 2M$ appearing in equation~(\ref{b-sph}) can be easily estimated using equation~(\ref{asymp}). One finds
\beq
\sum_{l=1}^{\infty}(2l + 1)\,P_{l}(\hat{r} \cdot \hat{r}')\overleftarrow{R}^{(1)}_{\omega l}(r)\overleftarrow{R}^{(1 *)}_{\omega l}(r') \approx \sum_{l=1}^{\infty}\frac{l(l+1)(2l+1)P_{l}(\hat{r} \cdot \hat{r}')|{\cal B}_{l}(\omega)|^2 e^{-i\omega (r_{*} - r'_{*})}}{(2M)^4\omega^2},\,\,\, r, r' \to 2M.
\label{asymp-2m3}
\eeq
For ${\bf x} = {\bf x}'$:
\beq
\sum_{l=1}^{\infty} (2l + 1)|\overleftarrow{R}_{\omega l}|^2 \approx \sum_{l=1}^{\infty}\frac{l(l+1)(2l+1)|{\cal B}_{l}(\omega)|^2}{(2M)^4\omega^2},\,\,\, r \to 2M.
\label{asymp-2m4}
\eeq
Other important estimate that one may evaluate is the one in which, say, $r \to \infty$ but $r' \to 2M$. One finds that
\bea
&&\sum_{l=1}^{\infty}(2l + 1)\,P_{l}(\hat{r} \cdot \hat{r}')\overleftarrow{R}^{(1)}_{\omega l}(r)\overleftarrow{R}^{(1 *)}_{\omega l}(r') 
\nn\\
&&\approx \sum_{l=1}^{\infty}\frac{l(l+1) (2l + 1)\,P_{l}(\hat{r} \cdot \hat{r}'){\cal B}^{*}_{l}(\omega)}{(2M)^2 r^2  \omega^2}\left(e^{-i\omega (r_{*} - r'_{*})} + \overleftarrow{{\cal A}}_{l}(\omega)e^{i\omega (r_{*} + r'_{*})}\right),\,\,\, r \to \infty, r' \to 2M.
\nn\\
\label{asymp-inf-2m}
\eea
The estimate for the other mode sum yields
\bea
&&\sum_{l=1}^{\infty}(2l + 1)\,P_{l}(\hat{r} \cdot \hat{r}')\overrightarrow{R}^{(1)}_{\omega l}(r)\overrightarrow{R}^{(1 *)}_{\omega l}(r') 
\nn\\
&&\approx \frac{2e^{-i\xi/2}}{\Gamma(i\xi)}\sum_{l=1}^{\infty}\frac {l(l+1) (2l + 1)\,P_{l}(\hat{r} \cdot \hat{r}'){\cal B}_{l}(\omega) e^{i\omega r_{*}} e^{i\xi\ln l}}{(2M)^2 r^2  \omega^2}K_{4M\omega i}\left(2l\sqrt{g_{00}(r')}\right),\,\,\, r \to \infty, r' \to 2M.
\nn\\
\label{asymp-inf-2m2}
\eea
Observe that such expressions yield a vanishingly small contribution to equation~(\ref{b-sph}) as $r \to \infty$ and thence can be neglected. For a fixed $r$ and $r' \to 2M$ one gets
\bea
&&\sum_{l=1}^{\infty}(2l + 1)\,P_{l}(\hat{r} \cdot \hat{r}')\overrightarrow{R}^{(1)}_{\omega l}(r)\overrightarrow{R}^{(1 *)}_{\omega l}(r') 
\nn\\
&&\approx \frac{2e^{-i\xi/2}}{\Gamma(i\xi)}\sum_{l=1}^{\infty}\frac {l(l+1) (2l + 1)\,P_{l}(\hat{r} \cdot \hat{r}')\overrightarrow{{\cal R}}^{(1)}_{\omega l}(r) e^{i\xi\ln l}}{(2M)^2 r^2  \omega^2}K_{4M\omega i}\left(2l\sqrt{g_{00}(r')}\right),
\nn\\
\label{asymp-2m6}
\eea
and
\beq
\sum_{l=1}^{\infty}(2l + 1)\,P_{l}(\hat{r} \cdot \hat{r}')\overleftarrow{R}^{(1)}_{\omega l}(r)\overleftarrow{R}^{(1 *)}_{\omega l}(r') 
\approx \sum_{l=1}^{\infty}\frac{l(l+1) (2l + 1)\,P_{l}(\hat{r} \cdot \hat{r}')\overleftarrow{{\cal R}}^{(1)}_{\omega l}(r){\cal B}^{*}_{l}(\omega) e^{i\omega  r'_{*}}}{(2M)^2 r^2  \omega^2}.
\label{asymp-2m5}
\eeq
For other details concerning the evaluation of asymptotic correlation functions at equal space-time points, see Refs.~\cite{candelas,china}.

\end{widetext}

\end{document}